 \documentclass[journal, letterpaper]{IEEEtran}

\normalsize

\ifCLASSINFOpdf
   \usepackage[pdftex]{graphicx}
   \DeclareGraphicsExtensions{.pdf,.jpeg,.png}
\else
   \usepackage[dvips]{graphicx}
   \DeclareGraphicsExtensions{.eps}
\fi

\usepackage{verbatim} 
\usepackage[cmex10]{amsmath}
\usepackage{amssymb,amsfonts}
\usepackage{mathtools}  
\usepackage{wasysym}
\usepackage{bbm}
\usepackage{cases}
\usepackage{empheq}
\usepackage{color,soul}
\usepackage{color,xcolor}

\usepackage{theorem} 
\def\myproof{\noindent{{\textbf{Proof:}}}} 

\ifdefined\myproof
\else
\def\myproof{\proof}

\fi

\newtheorem{theorem}{Theorem}
\newtheorem{definition}{Definition}
\newtheorem{lemma}{Lemma}
\newtheorem{proposition}{Proposition}

\newtheorem{remark}{Remark}

\newtheorem{proof}{Proof} 

\usepackage{cite} 

\usepackage[tight,footnotesize]{subfigure}

\usepackage[linesnumbered,ruled]{algorithm2e}
\usepackage{multirow}

\begin{document}

%
\title{Fast-Fading Channel and Power  Optimization of the Magnetic Inductive Cellular Network}
%
%
%
%

%
%

\author{Honglei~Ma,
        Erwu~Liu*, ~\IEEEmembership{Senior Member,~IEEE,}
        Zhijun~Fang*, ~\IEEEmembership{Senior Member,~IEEE,}
        Rui~Wang, ~\IEEEmembership{Senior Member,~IEEE,}
        Yongbin~Gao,
        Wenjun~Yu,
        Dongming~Zhang
\thanks{This work was supported in part by National Natural Science Foundation of China under Grant 42171404,  in part by  Shanghai Local Capacity Enhancement project (No. 21010501500), in part by "Science and Technology Innovation Action Plan" of Shanghai Science and Technology Commission for social development project under Grant 21DZ1204900, in part by Interdisciplinary Key Project of Tongji University under Grants 2023-2-ZD-04; (\emph{Corresponding authors: Erwu Liu; Zhijun Fang})}
\thanks{ Honglei Ma, Zhijun Fang, Yongbin Gao, Wenjun Yu  are with the
  School of Electronic and Electrical Engineering, Shanghai University of Engineering Science, Shanghai, China, E-mail: holyma@yeah.net,   Zjfang@gmail.com, gaoyongbin@sues.edu.cn, yuwenjun@sues.edu.cn }
\thanks{ Erwu Liu, Rui Wang are with the School of
School Electronics and Information Engineering, Tongji University, Shanghai, China, E-mail: erwu.liu@ieee.org, ruiwang@tongji.edu.cn. }
\thanks{Dongming Zhang is with the College of Civil Engineering, Tongji University, Shanghai, China, E-mail: 09zhang@tongji.edu.cn. }
\thanks{This work has been accepted  by the IEEE TWC for  publication. The  DOI number is 10.1109/TWC.2024.3425473}
}

\markboth{}%
{Submitted paper}


\markboth{Accepted by IEEE Transactions On Wireless Communications in July 2   ~2024 }
{Shell \MakeLowercase{\textit{et al.}}: Fast-Fading Channel and Power  Optimization of the Magnetic Inductive Cellular Network}

\IEEEpubid{0000--0000/00\$00.00~\copyright~2021 IEEE}

%



\maketitle

\begin{abstract}
The cellular network of magnetic Induction (MI) communication   holds promise in long-distance underground environments. In the traditional MI communication, there is no  fast-fading channel since the MI channel is treated as a quasi-static channel. However, for the vehicle (mobile) MI (VMI) communication, the unpredictable antenna vibration  brings the remarkable fast-fading. As such fast-fading  cannot be modeled by the central limit theorem, it  differs radically from other wireless fast-fading channels. Unfortunately, few studies focus on this phenomenon. In this paper, using a novel space modeling based on the electromagnetic field theorem, we propose a 3-dimension model of the VMI antenna vibration. By proposing ``conjugate pseudo-piecewise functions'' and boundary $p(x)$ distribution, we derive the cumulative distribution function (CDF), probability density function (PDF) and the expectation  of the VMI fast-fading channel. We also theoretically analyze the effects of the VMI fast-fading on the network throughput, including the VMI outage probability which can be ignored in the traditional MI channel study. We draw several intriguing conclusions different from those in wireless fast-fading studies. For instance, the fast-fading brings more uniformly distributed channel coefficients.  Finally, we propose the power control algorithm  using the non-cooperative game and multiagent Q-learning methods to optimize the throughput of the cellular VMI network. Simulations validate the derivation and the proposed algorithm.
\end{abstract}


%
\IEEEpeerreviewmaketitle

\vspace{-0.0em}
\section{Introduction}\label{sect_intro}
%
%
%
%

\IEEEPARstart{W}{ith} the expansion of human activity into the underground space, communication capabilities are needed in many underground scenarios such as  underground coal mines,  subways and  tunnels. However, electromagnetic  wave (EMW) dramatically attenuates through the underground obstacles. Wired communication is costly and hard to be deployed in the emergency rescue scenarios and remote mountains. MI communication (MIC) has been proven to be a reliable communication technique for the underground environment\cite{Kisseleff2018Survey, Li2019Survey, Guo2017Practical,Chen2023Novel}. The researches of MIC extend to the study of the physical layer (e.g., the channel models\cite{Guo2017Practical,Sun2010Magnetic}, transmit rate\cite{kisseleff2013channel, sun2012capacity}, modulation\cite{Kisseleff2014Modulation}, channel estimation\cite{Kisseleff2014Transmitter, Tan2017environment},   MI multiple-input and multiple-output (MIMO)\cite{Wang2022Multi,Wang2023Backscatter} and power control\cite{lin2015distributed ,Guo2021Joint}),  the data link layer\cite{Chen2023Novel} and the network layer\cite{lin2015distributed}.
\IEEEpubidadjcol

For the long-distance MIC, researchers analyze the MIC with a stationary antenna and propose an MI waveguide method by deploying several passive relays between two transceivers \cite{Sun2010Magnetic}. Such method achieves a transmitting distance of 250m. However, for the vehicle MI communications (VMIC), it is infeasible to deploy the MI waveguide due to limited  underground free space. Thus, some researchers and our team investigate the long-range MIC method  using the 1--10KHz very low frequency signals and a larger antenna (VLF-LA)\cite{zhang2014cooperative,Zhang2017Connectivity,Ma2019Effect, Ma2019Antenna}. Using VLF-LA, we achieve a 300-meters deep MIC distance in a coal mine\cite{zhang2014cooperative},\cite{Zhang2017Connectivity}.  However, two issues should be considered: orientation sensitivity and spectrum shortage. Firstly, the VLF-LA based MI channel is significant orientation sensitivity due to the large transmit antenna with  the 1.2--8 m diameters \cite{zhang2014cooperative, Zhang2017Connectivity,Ma2019Effect,Ma2019Antenna, Ma2020Channel}. Although the MIMO antenna mentioned in\cite{Guo2017Practical,Dumphart2016Stochastic,Zhang2022Performance,Wang2022Multi,Li2022Optimal} can alleviate the orientation sensitivity,  it is difficult to apply such large  multi-antenna transmitter in vehicles and tunnels due to the limited space. Secondly, there are very limited spectrum resources  in the MIC. According to\cite{Ma2019Effect}, the VLF-LA method only achieves   a frequency bandwidth of 300--500Hz.

\begin{table*}[t!]  
\caption{Comparison among different models of the fast-fading channel in related references }
\centering
\begin{tabular}{|c|l|l|l|c|}
  \hline
  Types & Models & Compared points  & Different challenges of  VMIC (this paper)   & Refs.\\
  \hline \hline
  \centering
  \multirow{2}{*}{EMWC} & Rayleigh & \hspace{+0.0em}1) Under relatively predictable large-scale fading &  \hspace{+0.0em} \maltese 1) Vibrating-oriented fading significantly depends & \\ 
                        & Rician &2) Received signals with many independent& on the vehicle under unpredictable velocity   & \\ 
                        & Nakagami &components and different phases which can be  &  \maltese 2) Receive signals with only one component   & \cite{Simon2005Digital,Tse2005Fundamentals} \\
                         &  & modeled as a Gaussian involved distribution via & which cannot be  simplified by  the central limit  & \\  
                          &  & central limit theorem 3) No boundary limits & theorem \maltese 3) With boundary limits & \\ 
  \hline
  \multirow{2}{*}{FSOC} & M\'{a}laga  & 1)  Be simplified to Rayleigh,Rician & Same as \maltese 1), \maltese 2), \maltese 3)  & \cite{Farid2012Diversity,jurado2011unifying,Wang2021Hovering}\\
                         & &and Nakagami models  2) No boundary limits & & \\ \cline{2-5}
                          &GML & 1) Linear propagation signals & 1) Unpredictable  propagation direction 2) Angle & \cite{Najafi2020Statistical} \\ 
                          &  &  2) No boundary limits & based input parameters  3) Significant sensitivity of& \\  
                          &  &  & antenna angle 4) With boundary limits & \\  \cline{2-5}
  \hline
  \multirow{2}{*}{MIC} & Traditional & No fast-fading & With fast-fading & \cite{Kisseleff2018Survey, Li2019Survey, Guo2017Practical,Chen2023Novel,Sun2010Magnetic,kisseleff2013channel, sun2012capacity,Kisseleff2014Modulation,Kisseleff2014Transmitter,Tan2017environment,Wang2022Multi,Wang2023Backscatter,lin2015distributed,Guo2021Joint,zhang2014cooperative,Zhang2017Connectivity,Ma2019Effect, Ma2019Antenna}\\ \cline{2-5}
                        &Underwater &  1) Antenna orientation with uniform distributions    & 1)Antenna orientation with complicated distributions  &\cite{Dumphart2016Stochastic} \\ 
                        & &  2) No boundary limits & 2) With boundary limits & \\ \cline{2-5}
                          & Our prior & 1) 2-D model with a simple antenna polarization  & 1) 3-D model with a complicated antenna & \\ 
                       &  & gain 2) Antenna orientation with BCS distribution  &polarization gain  2) Antenna orientation with the &  \cite{Ma2020Channel} \\ 
                         &  & 3)  Applying under the horizontal road scenarios  & general distribution and  BCS distribution &   \\ 
                        &  & &  3) Applying without any spatial constraints&   \\ 

  \hline
\end{tabular}
\label{tbl_diff}

\vspace{-1.2em}
  \end{table*}

Most  previous studies are done for the stationary MI antenna. The channel gain of the MI link is quasi-static. Thus, some  parameters (e.g., outage probability) widely used in EMW based communication (EMWC) are ignored or meaningless.
For the vehicle  antenna, since there is a random instantaneous orientation introduced by the MI antenna vibration, the mobile MI channel is not quasi-static.
In the first chapter of \cite{Tse2005Fundamentals}, the fast-fading channel is defined as the channel varies significantly over the time-scale of communication. According to previous works on the achievable rate of long-distance MICs using VLF-LA\cite{lin2015distributed, Zhang2017Connectivity, Ma2019Antenna}, the Shannon's capacity of the MIC channel is below 0.5--10 Kbps under a MIC distance of 50m. Notably, although the MIC device developed by our team achieves a text message transmission to the subsurface over 300-meters deep in a coal mine environment in Datong, China\cite{Zhang2017Connectivity}, the transmit rate of this device is much smaller than its Shannon's capacity. Due to the low transmit rate, the VMIC data applications have a laxer delay requirement, especially in the  frame-based systems similar to \cite{lin2015distributed}. According to ISO/TC108 and \cite{Sheng2012Vehicle}, the range of the frequency spectrum  of the road disturbance input is about 10--1000Hz. This means that the VMIC channel gain may change rapidly during a symbol time. The VMI channel contains the fast-fading gain.  With this MI fast-fading, parameters such as the outage probability become crucial to the performance of the MIC.

The MI fast-fading gain  differs radically from that of EMWC  and other communications (see Table \ref{tbl_diff}). In EMW-based channels, fast-fading gain is primarily caused by signal propagation, such as multipath effects  \cite{Tse2005Fundamentals}. Besides the multipath-based fading, the fluctuation-based fading is also a fast-fading occurring in the  free-space optical (FSO) communication (FSOC)\cite{Farid2012Diversity,jurado2011unifying} and unmanned aerial vehicle (UAV) based communication systems\cite{Wang2021Hovering, Najafi2020Statistical}. In \cite{jurado2011unifying}, the FSO link is modeled by a M\'{a}laga fading channel with three typical turbulent conditions, including irradiance fluctuation, line-of-sight contribution and scattering. Specifically, in the  UAV-assisted system, the geometric and misalignment losses (GML) are considered for the fast-fading modeling.  However, in the mobile MIC system,  antenna vibration causes the antenna (coil) misalignment, which predominantly contributes to the  fast-fading gain. Thus,  contrary to the EWMC and FSOC, the central limit theorem is not applicable for modeling the MIC channel.   Unlike the linear propagation of EMW and optical signals, MI signals are transmitted through magnetic field curves, making it impractical to determine signal propagation direction and align antennas for deployment. Even slight variations in magnetic field curve orientation and antenna alignment can lead to significant changes (even to zero) in the MIC channel power gain, as illustrated in Fig. 5 of \cite{Ma2019Effect}.
In addition, the MI antenna vibration is more unpredictable since it depends on the antenna carrier, road roughness and  vehicle velocity (subjective actors of drivers).

For the stochastic misalignment of MI antennas, in\cite{Dumphart2016Stochastic} and \cite{Zhang2022Performance},  researchers investigate the stochastic misalignment of antennas for underwater which are the precursor models of the MI fast-fading gain. In \cite{Ma2020Channel}, researchers propose  the concept of MI fast-fading  for the first time. In\cite{Dumphart2016Stochastic} and \cite{Zhang2022Performance}, researchers assume that the antenna orientation obeys the uniform distribution. Compared to  underwater scenarios,  the antenna orientation does not follow the uniform process in underground scenarios. Instead, it may follow a complex random process\cite{Tuerkay2005A}. In \cite{Ma2020Channel}, researchers  study the VMI fast-fading based on the 2-dimension (2-D)  underground scenarios. Except \cite{Dumphart2016Stochastic,Zhang2022Performance, Ma2020Channel}, there seems to be no other research discussing the random antenna orientation. Furthermore, unlike in free space, as the MI antenna is a coil deployed on its carrier, it precludes unrestricted rotation. It is imperative to consider the mechanical or electronic limits of  antenna vibration, called vibration boundaries.

For the spectrum shortage issue,  the  widely-used  cellular topology is a cost-effective option.   But it is desirable to design the optimal allocation strategy to give full play to the advantages of cellular network\cite{Li2022Dynamic}.  In the MIC study,  aiming at the higher overall MI network throughput,  the study \cite{lin2015distributed} uses the distributed power control optimization (PCO) method with the Nash equilibrium (NE) game.  As the vehicle velocity and the road roughness are more unpredictable than the propagation of EMW, the MI fast-fading gain may  change  more dramatically  than  EMW fast-fading gain during tens of seconds.  The  NE game  as the one in \cite{lin2015distributed} is not suitable for solving the PCO problem with such frequently changing environment. Therefore, like \cite{Chen2013stochastic},\cite{Bennis2013self}, we introduce a non-cooperative reinforcement learning (RL) to keep  the optimal power strategy learned in real time. Also, due to the extremely low bandwidth of the long-distance MI link, the deep RL methods  as those  in\cite{Nasir2019Multiagent},\cite{Wang2022Decentralized} with the online training process are not suitable for solving the MI power control problem.

In this work, we derive the closed-form expressions about the  statistical characteristics  of the VMI fast-fading gain. Based on these expressions, we propose a  distributed PCO algorithm  to compute power allocations in real time for every BS of  the  cellular-topology based vehicle MICs (CVMC). The statistical characteristics  of the VMI fast-fading channel serve as  the foundation for further related studies (Table \ref{tbl_contr}).

  \begin{table}[t!]  
\caption{Potential further  contributions to the MIC fields }
\centering
  \begin{tabular}{|c|c|}
      \hline
     Typical fields of MICs & Typical refs. \\
        \hline \hline
      Channel capacity and data rate maximization & \cite{sun2012capacity,lin2015distributed}\\
       \hline
       Channel estimations & \cite{Kisseleff2014Transmitter, Tan2017environment} \\
       \hline
         Spatial diversity for MI networks or cooperative MICs & \cite{Ma2019Effect,Ma2019Antenna} \\
       \hline
                 Transmission range Extension&  \cite{zhang2014cooperative}\\  
      \hline
       Connectivity of networks &  \cite{Zhang2017Connectivity}\\
      \hline
            Antenna orientation optimizations&  \cite{zhang2014cooperative,Ma2019Antenna} \\  
        \hline
                 Error control&  \cite{lin2015distributed}\\  
      \hline
               Cross-layer solutions such as network protocol& \cite{Kisseleff2018Survey}  \\
        design, interference manage, cross-layer optimization & \cite{lin2015distributed, Kisseleff2014Modulation} \\
      \hline
    \end{tabular}
    \label{tbl_contr}
  \end{table}

Different from the works \cite{Dumphart2016Stochastic} and \cite{Zhang2022Performance} for the underwater scenarios, the VMI antenna orientation in this paper does not obey the uniform distribution. Also, they  do not consider the  boundary events of the antenna.  Different from the works \cite{Ma2020Channel} with 2-D antenna vibration models, the 3-D model of the antenna vibration contains several dependent random components on the antenna orientation.  We need to transform them to be independent first. For the PCO, different from the work\cite{lin2015distributed} in the MIC area, since the VMI channel is not quasi-static, the PCO aiming at the instantaneous throughput  is not suitable for VMI PCO problem. To sum up, we make the following contributions.
 \begin{itemize}
   \item {Using a novel space modeling, we propose a 3-D model of the VMI antenna vibration. Using such model, we obtain an function of VMI fast-fading gain with a single random variable. We introduce a novel definition of ``conjugate pseudo-piecewise functions" to refine the monotonicity intervals of the VMI fast-fading gain.  To the best of  our knowledge, this  paper is the first to consider the 3-D fast-fading phenomenon for the long-distance  underground mobile MIC. }
   \item {After defining  the  boundary $p(x)$ distribution, we derive the closed-form expressions of the CDF, PDF and expectation of the fast-fading gain of  the  CVMC under the  general distribution vibration model. We use the Dirac's function to model the boundary effects for a random variable. Specifically, the  Gaussian road roughness model is substituted into this general model, yielding several intriguing conclusions   distinct from those in the EMWC base communications. For instance, the fast-fading has the potential to enhance overall performance through a suitable larger antenna vibration intensity (AVI).
      }
   \item {We derive the closed-form expression of the outage probability of the CVMC. From this, we uncover an intriguing  phenomenon that the larger AVI might sometimes reduce the  outage probability of the download CVMC under the Gaussian road roughness model. }
    \item {We propose the distributed CVMC power control optimization (CVMC-PCO) algorithm with multiagent Q-learning and non-cooperative game methods to achieve the optimal throughput of the CVMC network.
      }
 \end{itemize}

 This paper is structured as follows. Section \ref{sect_vibration} presents the VMI  antenna vibration model and the VMI channel gain. Section \ref{sect_fastfading} proposes the  statistical characteristics  of the VMI fast-fading gain.  Section \ref{sect RL} discusses the PCO problem over multi-cells. Section \ref{sect sim} validates the findings. Section \ref{sect conclusions} concludes the paper.

\emph{ Notation conventions:} Hereafter, $(\cdot)_{b, i}$ denotes the variable $(\cdot)$ of the link  $b\rightarrow i$,  $(\cdot)_i$ denotes the variable $(\cdot)$ of  the $i$-th vehicle user (VU),  $(\cdot)_{[k, n]}$  represents the variable $(\cdot)$ of the $n$-th VU in the $k$-th cell/subnetwork. $\mathbb{E}$ denotes the expectation, $\mathbb{P}$ denotes the probability, $(\cdot)^{T}$ denotes transpose. Boldface capital letters, such as $\mathbf{H}$, denote  matrices or vectors.   Boldface lowercase characters,  such as $\mathbf{n}$, denote vectors. The coil serves as the MIC antenna.

\section{Antenna Vibration Modeling }\label{sect_vibration}

In this section, we propose a model of an antenna vibration and extract the independent VMI fast-fading parameters from the VMI channel power gain. By constructing so-called ``pseudo-piecewise functions'',  we derive two lemmas regarding the geometric characteristics of the  fast-fading gain  for the derivations of statistic characteristics.

The MIC is realized by using a couple of coils. In the 3-D space, its channel gain is determined by the mutual induction $M_{tr}$ between a transmitter coil and a receiver coil
\begin{equation}\label{eqn_mtr}
\begin{aligned}
M_{tr} = M_0 (\cos\theta_{\mathrm{tx}} \cos\theta_{\mathrm{rx}} - \frac{1}{2} \sin \theta_{\mathrm{tx}} \sin\theta_{\mathrm{rx}}), \notag
\end{aligned}
\end{equation}
where $\theta_{\mathrm{tx}}$ and $\theta_{\mathrm{rx}}$ are the angles between the coil radial directions and the line connecting the two coil centers, respectively\cite{Tan2015Environment}, $M_0$ denotes the   mutual inductance  between two aligned coils. However, $\theta_{\mathrm{tx}}$ and $\theta_{\mathrm{tx}}$ depend on each other even when the transmit coil remains horizontal.
 Therefore, two challenging principles are considered in antenna vibration modeling: 1) to minimize the number of random variables as much as possible,  and 2) to reduce the dependencies of  random variables as much as possible.

 \vspace{-0.81em}
\subsection{The Modeling}
\begin{figure}[t]
        \centering
        \includegraphics[width=3.2in]{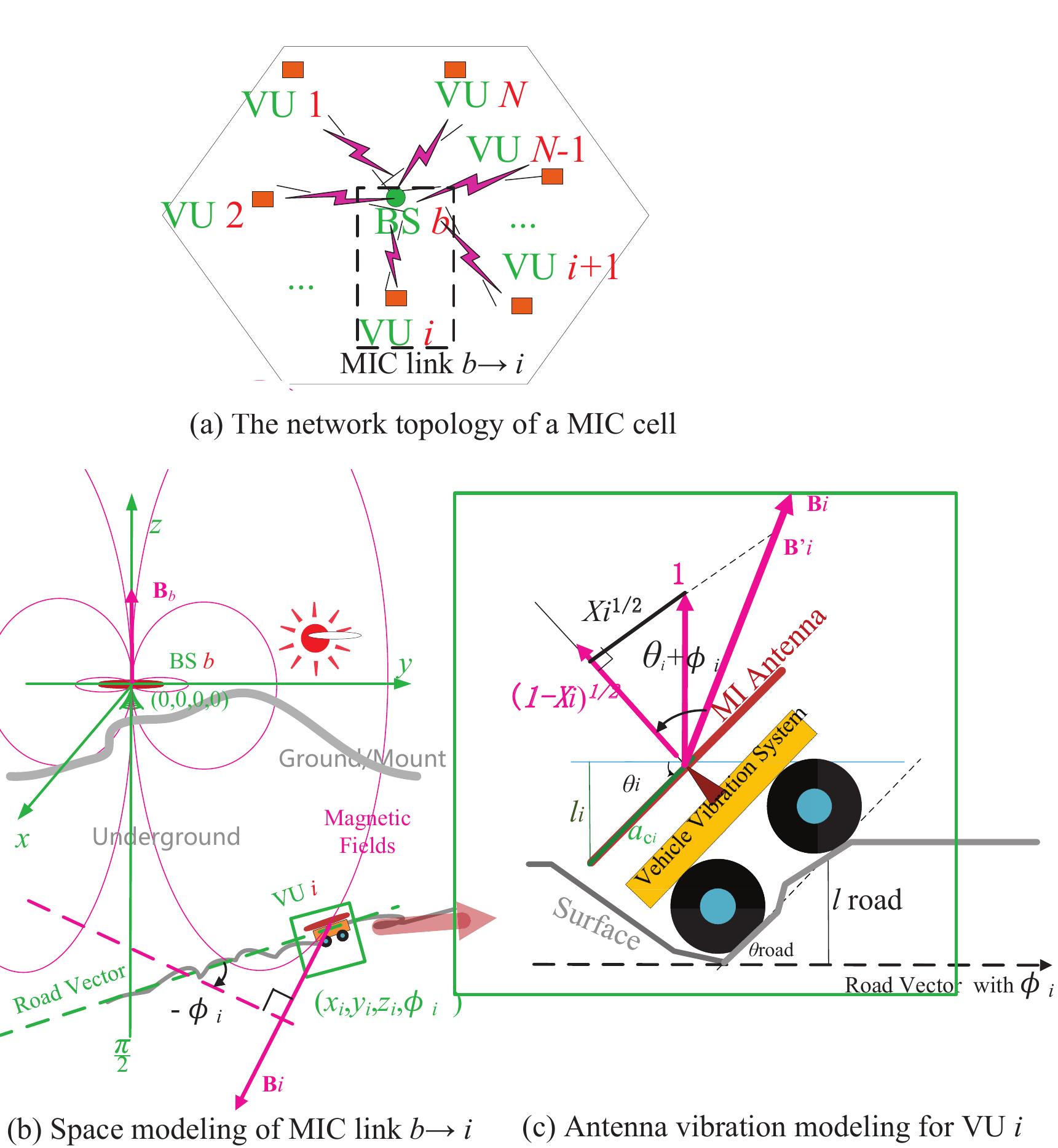}\\
        \vspace{-0.5em}
        \caption{Antenna vibration modeling for  the channel from BS $b$ to VU $i$,  $\theta_i \in [-90^\circ,90^\circ]$ is the angle of the antenna vibration, $\phi\in[-180^\circ, 180^\circ]$ represents the antenna background orientation, $X_i$ is the AVI.  
 }\label{fig_download}
 \vspace{-0.401em}
    \end{figure}

\vspace{-0.0em}

\begin{figure}[t]
        \centering
        \includegraphics[width=1.8in]{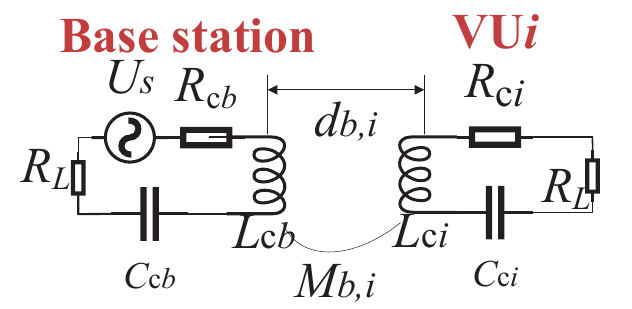}\\
        \vspace{-0.5em}
        \caption{ Equivalent circuit model for   the channel from BS to VU $i$.  
 }\label{fig_circuit}
 \vspace{-0.01em}
    \end{figure}

As shown in Fig. \ref{fig_download}(a),  we assume the existence of a star topology network within a MIC cell for the underground communication, where the $N$ VUs and the base station (BS) $b$ form $N$ MIC links.  For any MIC link $b\rightarrow i$, we assume the road adjacent to VU $i$ is sloped and rough, with potholes of varying sizes that are sufficiently dense and uniformly distributed. In this paper, we postulate the vehicle and its antenna system to be ideal. That is to say,  the antenna center is situated precisely atop the centroid of the vehicle.
For such road scenarios and an ideal vehicle system, the likelihood of vehicle vibration occurring in any horizontal direction is equivalent. Thus, if we observe  the plane of the coil from a single horizontal perspective, we propose a 3-D model of  antenna vibration which can be simplified to a single independent antenna orientation component as depicted in Fig. \ref{fig_download}(b) and Fig. \ref{fig_download}(c).

In Fig. \ref{fig_download}(b) and Fig. \ref{fig_download}(c), we assume that the MI  BS is located at  $O (0, 0, 0)$ with horizontal circular antenna. The VU $i$   drives  on the sloped and bumpy underground road. At time $t$, the  VU $i$ located at  $(x_i, y_i, z_i$) where the angle between the  road  and the magnetic field is $\phi_i$. 

The MIC is realized by using a couple of coils. A modulated sinusoidal current in the transmitting coil will induce around it a time varying magnetic field in space, which further induces a sinusoidal current in the receiving coil. This current can be demodulated to information\cite{Zhang2017Connectivity}. Thus, a single link from a BS (transmitter) coil $b$ to a VU (receiving) coil $i$ is shown in Fig. \ref{fig_circuit}.
 Namely, the BS contains a coil with  $N_{\mathrm{c}b}$ turns and $a_{\mathrm{c}b}$ radius. The $i$-th VU node contains a coil with $N_{\mathrm{c}i}$ turns and $a_{\mathrm{c}i}$ radius. Each coil is made of a 17-AWG wire whose unit length resistance is $\rho_{w}$. Hence, the resistors of the coils of the BS and $i$-th VU are $R_{\mathrm{c}b} = 2N_{\mathrm{c}b}   \pi a_{\mathrm{c}b} \rho_w$ and $R_{\mathrm{c}i} = 2N_{\mathrm{c}i}   \pi a_{\mathrm{c}i} \rho_w$, respectively. The inductances  of the coilsof the BS and $i$-th VU are $L_{\mathrm{c}b} = \frac{1}{2} \pi  N_{\mathrm{c}b}^2  a_{\mathrm{c}b} \mu $ and $L_{\mathrm{c}i} = \frac{1}{2} \pi  N_{\mathrm{c}i}^2  a_{\mathrm{c}i} \mu $, respectively, where $\mu$ is the permeability of the space. $R_L$ is the load impendence.  We use the capacitors with capacitances $C_{\mathrm{c}b}=\frac{1}{(2\pi f_0)^2L_{\mathrm{c}{b}}}$ and $C_{\mathrm{c}i}=\frac{1}{(2\pi f_0)^2L_{\mathrm{c}{i}}}$ to tune the circuit resonances of the BS and $i$-th VU at a target resonance frequency $f_0$\cite{Ma2019Effect}, respectively.
 
As depicted in Fig. \ref{fig_download}(c), at time $t+\tau$ where $\tau$ is extremely small, an antenna vibration event occurs from a pothole. This pothole creates a random height difference $l_{\mathrm{road}}$ between the two vehicle wheels. Here, $l_i = \mathcal{V}(l_{\rm road})$ where $\mathcal{V}(\cdot)$ denotes a vehicle system leading to an   additional random deviation in the angle $\theta_i$ from $\phi_i$. Thus, at this time, the angle between the normal vector of the  coil of $i$-th VU  and the magnetic flux density($\mathbf{B}_i$) is $(\theta_i + \phi_i)$, called antenna polarization angle.


\vspace{-0.5em}

\subsection{MI  Fast-Fading Gain}

In this subsection, we analyze  the antenna vibration component ($\theta_i$) and the VMI fast-fading gain. Also, we reduce monotonicity intervals of the VMI fast-fading gain by the ``conjugate pseudo-piecewise function".

In \cite{Ma2019Effect}, we  derive the expression of  magnetic field  at Cartesian coordinate $(\zeta_i, z_i)$ in the 2-D plane:
\begin{equation}\label{eqn_Hixy}
\begin{aligned}
\mathbf{H}_{b,i} &= \frac{N_{\mathrm{c}b} I_0 a_{\mathrm{c}b}^2}{4 d^3_{b,i}e^{\frac{d_{b,i}}{\delta_{\rm co}}}} \left( \frac{2z_{i}^2 - \zeta_{i}^2}{\zeta_{i}^2+z^2_{i}}, \frac{3 \zeta_{i}z_{i} }{\zeta_{i}^2+z_{i}^2}\right)^T,
\end{aligned} 
\vspace{-0.2em}
\end{equation}
where $d_{b,i}=\sqrt{\zeta_{i}^2 + z_i^2}$  is the distance  between the transmitter and receiver. We transform~\eqref{eqn_Hixy} to the 3-D based expression by $\zeta_i=\sqrt{x_i^2+y_i^2}$. The mutual inductance is  the magnetic flux passing through the coil of $i$-th VU, i.e.,
\begin{equation}\label{eqn_mbiphi0}
\begin{aligned}
&M_{b,i} =  \frac{ \pi\mu a^2_{\mathrm{c}i}(\mathbf{H}_{b,i} \cdot \mathbf{n}_i)}{I_{\mathrm{co}}} \!=\!  \frac{ \pi\mu a^2_{\mathrm{c}i}||\mathbf{H}_{b,i}||\cos(\theta_i\! +\! \phi_i)}{I_{\mathrm{co}}}\\
         &\ \ = \frac{\pi\mu N_{\mathrm{c}i} N_{\mathrm{c}b}a^2_{\mathrm{c}i} a^2_{\mathrm{c}b}}{4 }  \exp(-\frac{\sqrt{x_i^2 + y_i^2 + z_i^2}}{\delta_{\mathbf{e}}})\\
         & \    \sqrt{\frac{(2z_{i}^2 - x_{i}^2 - y_i^2)^2 +  9(x_i^2+y_i^2) z_{i}^2  }{(x_i^2 + y_i^2 + z_i^2)^5}}
          \cos (\theta_i + \phi_{i}),
\end{aligned}
\vspace{-0.2em}
\end{equation}
where   $I_0$ is transmitting current of the BS antenna, $\mathbf{n}_i$ is the normal vector of the coil of  VU  $i$,  $\delta_\mathbf{e}$ is the skin depth. Here, for the convenience of separating vibration components, the inner product $(\mathbf{H}_i \cdot \mathbf{n}_i)$ can be expressed by ($||\mathbf{H}_i||\cos(\theta_i\! +\! \phi_i)$).
Let
\begin{equation}\label{eqn_mbimax}
\begin{aligned}
 \mathcal{M}_{b,i} &= \frac{\pi\mu N_{\mathrm{c}i} N_{\mathrm{c}b}a^2_{\mathrm{c}i} a^2_{\mathrm{c}b}}{4}  \exp(-\frac{\sqrt{x_i^2 + y_i^2 + z_i^2}}{\delta_{\mathbf{e}}})\\
         &  \sqrt{\frac{(2z_{i}^2 - x_{i}^2 - y_i^2)^2 +  9(x_i^2+y_i^2) z_{i}^2  }{(x_i^2 + y_i^2 + z_i^2)^5}}.
\end{aligned}
\vspace{-0.2em}
\end{equation}
With $\cos (\theta_i + \phi_i) = \cos\theta_i \cos\phi_i - \sin\theta_i\sin\phi_i$, we have
\begin{equation}\label{eqn_mbiphi1}
\begin{aligned}
M_{b,i} &=  \mathcal{M}_{b,i} \cos\phi_i\cos \theta_i - \mathcal{M}_{b,i}\sin\phi_i\sin\theta_i.
\end{aligned}
\vspace{-0.2em}
\end{equation}
According to \cite{Ma2019Antenna}, the channel power gain of the link BS$\rightarrow$VU $i$ is
\begin{equation}\label{eqn_Gbi0}
\begin{aligned}
G_{b,i} &= \left|\tfrac{(2\pi f)^2 R_L}{ Z_{\rm rx}^2 Z_{\rm tx}}\right| M^2_{b,i}  \\
&=   \left|\tfrac{(2\pi f)^2 R_L}{ Z_{\rm rx}^2 Z_{\rm tx}}\right| \mathcal{M}^2_{b,i}\left(\cos\theta_i\cos\phi_i - \sin\theta_i\sin\phi_i\right)^2,
\end{aligned}
\end{equation}
where  $Z_\text{tx}$$=$$j 2\pi f {L_{\mathrm{c}b}}$$+$$\frac{1}{j2\pi f C_{\mathrm{c}b}}$$+$$R_{\mathrm{c}b}$$+$$R_L$ and  $Z_\text{rx}$$=$$j 2\pi f {L_{\mathrm{c}i}}$$+$$\frac{1}{j2\pi f C_{\mathrm{c}i}}$$+$$R_{\mathrm{c}i}$$+$$R_L$ denote the overall circuit impedances {of the antennas of the transmitter and receiver, respectively.  We divide $G_{b,i}$ into three factors:  $\mathcal{G}_{\rm co}=\left|\tfrac{(2\pi f)^2 R_L}{ Z_{\rm rx}^2 Z_{\rm tx}}\right|$, $\mathcal{M}^2_{b,i}$ and 
\begin{equation}\label{eqn_Jbi00}
\begin{aligned}                  
J_{b,i} &= J^\diamond_{b,i}(Y_i) =  \left(\cos\theta_i\cos\phi_i - \sin\theta_i\sin\phi_i\right)^2.
\end{aligned}
\end{equation}
The factor $\mathcal{G}_{\rm co}$ is determined by   the electrical characteristics of  BS and $i$-th VU antennas. The factor  $\mathcal{M}^2_{b,i}$ depends on the location of the $i$-th VU antenna ($x_i,y_i, z_i$). Their product $\mathcal{G}_i = \mathcal{G}_{\rm co}\mathcal{M}^2_i$ indicates the channel power gain  between the aligned coils of transmitter and receiver (i.e., $\theta_{i}$$+$$\phi_{i}$$=$$0$). Such gain is called  \emph{MI aligned gain} which also implies the maximum energy  received at the location $(x_i, y_i, z_i)$.  Obviously, the MI aligned gain can be pre-observed. The factor $J_{b,i}$,  called \emph{MI polarization gain},  is determined by the polarization angle ($\phi_i + \theta_i$). To accurately calculate the angle $\phi_i$, we divide $\phi_i$ into the road  gradient $\phi_{\mathrm{road}i}$ and the angle $\phi_{\mathrm{p}i}$. The angle $\phi_{\mathrm{p}i}$ denotes the angle between the magnetic field and z-axis. Thus, $\phi_i =\phi_{\mathrm{p}i} + \phi_{\mathrm{road}i}$ holds. In general, the road  gradient $\phi_{\mathrm{road}i}$ is observed by tools. The variable $\phi_{\mathrm{p}i}$ is determined by the VU position and can be derived from~\eqref{eqn_Hixy}, namely,  $\phi_{\mathrm{p}i}$ is the first item of
 \begin{equation}\label{eqn_mbiphi0sinsin}
\begin{aligned}
\phi_i &= \arccos\left( \frac{\frac{2z_{i}^2 - \zeta_{i}^2}{\zeta_{i}^2+z_{i}^2}}{  \sqrt{(\frac{2z_{i}^2 - \zeta_{i}^2}{\zeta_{i}^2+z_{i}^2})^2 +  (\frac{3 \zeta_{i}z_{i} }{\zeta_{i}^2+z_{i}^2})^2    }   } \right) \!+\! \phi_{\mathrm{road}i} \\
     &= \arccos\left( \frac{2z_{i}^2 \!-\! x_{i}^2 \!-\! y_i^2}{  \sqrt{(2z_{i}^2 \!-\! x_{i}^2 \!-\! y_i^2)^2 \!+\!  9(x_i^2+y_i^2) z_{i}^2     }   } \right) \!+\!\phi_{\mathrm{road}i},
\end{aligned}
\end{equation}
which indicates that  $\phi_i$ is also pre-observable. We called  this pre-observable  $\phi_i$ as the \emph{antenna background orientation} (ABO).

 Subsequently, we  focus on $\theta_i$.  From our antenna vibration model (Fig. \ref{fig_download}(c)), we observe that the angle $\theta_i$ satisfies $\sin\theta_i$$=$$  \frac{l_i}{a_{\mathrm{c}i}}$. Here, $l_i$ is the  height difference between the antenna edge and the antenna center (HDAEAC). When the vibration angle $\theta_i$ reaches its limits $\pm\frac{\pi}{2}$, $\sin\theta_i=\pm 1$ holds. Namely,  $|l_i| \leq a_{\mathrm{c}i}$ always holds.  Thus, $Y_i$$=$$\sin\theta_i$$\in$$[-1,1]$ can be treated as a normalized HDAEAC, called normalized vehicle vibration  offset (NVVO). Considering the data collection in practical applications, we introduce  the power of NVVO, denoted as $X_i$$\triangleq$$Y^2_i$, as a measure of the antenna vibration intensity (AVI). 

After we extend the right of~\eqref{eqn_Jbi00} and substitute $Y_i = \sin\theta_i$, $1-Y_i^2 = \cos^2\theta_i$ into \eqref{eqn_Jbi00}, we obtain
\begin{equation}\label{eqn_Jbi01}
\begin{aligned}                  
J_{b,i}\!= \!J^\diamond_{b,i}(Y_i) &=  \cos^2(\phi_i) (1 - Y_i^2) + \sin^2(\phi_i) Y_i^2    \\
       &- 2\cos(\phi_i)\sin(\phi_i)  Y_i \sqrt{1-Y_i^2}.
\end{aligned}
\end{equation}

{For the moving VU, as the random vibration $\theta_i$ ($Y_i$) fluctuates fast throughout the symbol duration,  $J_{b,i}=J^\diamond_{b,i}(Y_i)$ is the fast-fading, as defined in \cite{Tse2005Fundamentals}. We  define $J_{b,i}$ as \emph{VMI fast-fading gain}. As the derivations of the statistical characteristics are the basis for analyzing the throughput for VMI networks, they are the most important works in this study.

Unfortunately, as $J^\diamond_{b,i}(Y_i)$ is a multimodal function  without  exhibiting parity, we cannot directly derive the PDF of $J_{b,i}$ from the PDF of the  NVVO $Y_i$ or AVI $X_i$. Instead, we should first investigate the monotonicities of  $J_{b,i}$ as follows.

According to~\eqref{eqn_mbiphi0} and~\eqref{eqn_Jbi01}, the  MI fast-fading gain can be rewritten as $J_{b,i} = \cos^2(\theta_i \! + \! \phi_i) = \frac{1}{2}\cos2(\theta_i\!+\!\phi_i)+\frac{1}{2}$. Such cosine function indicates that the stationary points are at $\theta_i\!=\!k\pi\!-\!\phi_i$ and $\theta_i\!=\!\frac{\pi}{2}\!-\!\phi_i \!+\! k\pi$ ($k$ is an integer), respectively. Due to $\theta_i \in [-\frac{\pi}{2}, \frac{\pi}{2}]$, we obtain two stationary points
  \begin{equation}\label{eqn_mathcalXN}
  \begin{aligned}
 \mathcal{Y}^*_{i-}    = \!\begin{cases}
     -\sin\phi_i      \ \ \ \ \ \ \ \ &\phi_i \in (k\pi,k \pi + \frac{\pi}{2}), \\
     -\frac{1}{\sqrt{1+\tan^2\phi_i}}      \ \ \ \ \ \ \  \  &\phi_i \in (-k\pi -\frac{\pi}{2}, -k\pi),
  \end{cases}
  \end{aligned}
 \end{equation}
and
  \begin{equation}\label{eqn_mathcalXP}
  \begin{aligned}
 \mathcal{Y}^*_{i+}
    = \!\begin{cases}
     \frac{1}{\sqrt{1+\tan^2\phi_i}}      \ \ \ \ \ \ \ \ &\phi_i \in (k\pi,k \pi + \frac{\pi}{2}), \\
     -\sin\phi_i     \ \ \ \ \ \ \  \  &\phi_i \in (-k\pi -\frac{\pi}{2}, -k\pi),
  \end{cases}
  \end{aligned}
 \end{equation}
 respectively.  Therefore, the monotonicity of $J^\diamond_{b,i}(Y_i)$ be summarized as follows.
1) When $-1 \leq Y_i <  \mathcal{Y}^*_{i-}$, the function $J^\diamond_{b,i}(Y_i)$ is  monotonically increasing/decreasing. 2) When $\mathcal{Y}^*_{i-} \leq Y_i <  0$, the function $J^\diamond_{b,i}(Y_i)$ is  monotonically decreasing/increasing. 3)  When $0 \leq Y_i < \mathcal{Y}^*_{i+} $, the function $J^\diamond_{b,i}(Y_i)$ remains  monotonically decreasing/increasing. 4) When $\mathcal{Y}^*_{i+} \leq Y_i <  1$ the function $J^\diamond_{b,i}(Y_i)$ is  monotonically increasing/decreasing.

However, the monotonicity of $J^\diamond_{b,i}(Y_i)$  is determined by   $\mathcal{Y}^*_{i+}$, $\mathcal{Y}^*_{i-}$ and the signs of $Y_i$ and $\phi_i$. Moreover, there are  many forms of $J_{b,i}^{-1}(Y_i)$. Worse still, the preceding works may solely provide  statistical distributions and datasets of $Y_i^2$ or $|Y_i|$.   Thus, it remains complicated to study the statistical characteristics of the vehicle MI channel. To reduce  the number of the monotonicity intervals of $J_{b,i}$ for deriving its CDF, we propose the following definition.

\vspace{-0.1em}
\begin{definition}\label{def_piecewise}
Given a continuous domain  $\mathcal{D}$$\subset$$\Re^+$,   the operator $J^{\sqrt{.}}(\cdot)$ is said to be a \textbf{conjugate pseudo-piecewise function}, $\forall x $$\in$$ \mathcal{D}$,  if there  exists one variable  $x^{\sqrt{.}}$$\in$$-\mathcal{D}^\frac{1}{2}\bigcup\mathcal{D}^\frac{1}{2}$ chosen for matching  $(x^{\sqrt{.}})^2 $$=$$ x$ and $J^{\sqrt{.}}(\cdot)$ satisfies
\begin{equation}
\begin{aligned}
   J^{\sqrt{.}}(x) = \begin{cases}
      J_1(x)      &  \ x^{\sqrt{.}} < 0,  \\
      J_2(x)      &  \ x^{\sqrt{.}} \geq 0,  \\
  \end{cases}
\end{aligned}\notag
\end{equation}
when the $x^{\sqrt{.}}$ matches  $(x^{\sqrt{.}})^2 $$=$$ x$. 
\end{definition}

Obviously, for all AVIs $X_i$$\in$$ \mathcal{D}$$=$$[0,1]$, we can find at least one $Y_i\in -[\sqrt{0},\sqrt{1}]\bigcup[\sqrt{0},\sqrt{1}]$ matching $X_i$$=$$Y_i^2$. Since $Y_i$$=$$-\sqrt{X_i}$ when $Y_i$$<$$0$, 
we can rewrite the expression~\eqref{eqn_Jbi01} as the form of the conjugate pseudo-piecewise function
\begin{equation}\label{eqn_Jbi}
\begin{aligned}
  J_{b,i}\!=\!J_{b,i}^\diamond(Y_i) \!=\! J^{\sqrt{.}}_{b,i}(X_i) \!= \!\begin{cases}
      J_{b,i+}(X_i)      & Y_i \in [0,1], \\
      J_{b,i-}(X_i)      & Y_i \in [-1,0), \\
  \end{cases}
\end{aligned}
\end{equation}
where $X_i = Y^2_i$, 
\begin{equation}\label{eqn_Jip}
\begin{aligned}
  J_{b,i+}(X_i) &= \cos^2\phi_i (1 - |X_i|) + \sin^2\phi_i |X_i|     \\
       &- 2\cos\phi_i\sin\phi_i   \sqrt{|X_i|-|X_i|^2},
\end{aligned}
\end{equation}

\begin{equation}\label{eqn_Jin}
\begin{aligned}
  J_{b,i-}(X_i) &= \cos^2\phi_i (1 - |X_i|) + \sin^2\phi_i |X_i|    \\
       &+ 2\cos\phi_i\sin\phi_i   \sqrt{|X_i|-|X_i|^2}.
\end{aligned}
\end{equation}
Here, we say $Y_i$ is  an implicitly/matching independent variable with respect to $X_i$ for $J^{\sqrt{.}}_{b,i}(\cdot)$.

To reduce the number of the monotonicity discussions for the CDF derivation, we find the following lemmas for the conjugate pseudo-piecewise function
$J^{\sqrt{.}}_{b,i}(\cdot)$.
 \begin{lemma}\label{prop_jbiconvex}
(\textbf{Concavity}): The functions $J_{b,i+}(X_i)$ and $J_{b,i-}(X_i)$ are convex and concave for $\phi_i \in [0, \frac{\pi}{2}\!+\!k\pi]$, respectively. Conversely, the functions $J_{b,i+}(X_i)$ and $J_{b,i-}(X_i)$ are concave and convex for $\phi_i \in [-\frac{\pi}{2}\!-\!k\pi, 0]$, respectively.
\end{lemma}

\begin{myproof}
Please refer to Appendix \ref{sect_proofjbiconvex}.
\end{myproof}

 \begin{lemma}\label{prop_jbiXeven}
(\textbf{Duality}): If the AVI $X_i \in[0,1]$ is assumed as a  free variable, the functions $J_{b,i+}(X_i, \phi_i)$ and $J_{b,i-}(X_i, \phi_i)$ satisfy the following equations:
\begin{subequations}\label{eqn_jbiXeven}
\begin{align}
J_{b,i+}(X_i, \phi_i) &= J_{b,i-}(X_i, k\pi-\phi_i), \label{eqn_jbiXeven:1} \\
J_{b,i-}(X_i, \phi_i) &= J_{b,i+}(X_i, k\pi-\phi_i), \label{eqn_jbiXeven:2}\\
J_{b,i+}(0) &= J_{b,i-}(0) = \cos^2\phi_i, \label{eqn_jbiXeven:3}\\
J_{b,i-}(1) &= J_{b,i+}(1) = \sin^2\phi_i. \label{eqn_jbiXeven:4}
\end{align}
\end{subequations}
\end{lemma}

\begin{myproof}
After substituting $\cos(\phi_i) \!=\! \pm \cos( k\pi\!-\!\phi_i)$,  $\sin(\phi_i)\!=\!\pm \sin( k\pi\!-\!\phi_i)$,  $\sin(2\phi_i) \!=\! -\sin2( k\pi\!-\!\phi_i)$ and $2\sin\phi_i\cos\phi_i\!=\!\sin2\phi_i$ into~\eqref{eqn_Jin}, we can obtain~\eqref{eqn_jbiXeven:1},~\eqref{eqn_jbiXeven:2},~\eqref{eqn_jbiXeven:3} and~\eqref{eqn_jbiXeven:4}.

This concludes the proof.
\end{myproof}

Lemmas \ref{prop_jbiconvex}, \ref{prop_jbiXeven} indicate that  the curves of $J_{b,i+}(X_i)$  under  $\phi_i \in [0, \frac{\pi}{2}\!+\!k\pi]$ and $J_{b,i-}(X_i)$ under  $\phi_i \in  [-\frac{\pi}{2}\!-\!k\pi,0]$ overlap. This means that the number of $J_{b,i}$ monotonicities  halved. Additionally, it is observed from~\eqref{eqn_jbiXeven:3} and~\eqref{eqn_jbiXeven:4} that  $J_{b,i+}(X_i)$ and  $J_{b,i-}(X_i)$ combine into a closed curve.  Moreover, these  lemmas imply that there are only two different forms of real solutions $(J^{\sqrt{.}}_{b,i}(X_i)=0)$.  Also, $J_{b,i}$ is  independent of the signs of $\phi_i$ and $Y_i$. Thus, we can greatly streamline  the statistical characteristic models of the VMI fast-fading gain.

\section{Statistical Characteristics of VMI Fast-fading Gain}\label{sect_fastfading}


In this section, we derive the CDF, PDF and expectation of VMI fast-fading gain under two typical distribution vibration models. Then, we study the outage probability caused by the VMI fast-fading gain.

 We assume that the center of the antenna is directly above the center of mass of the vehicle. The NVVO has an even PDF $f_{Y_i}(y)$.

\subsection{ General  Distribution Vibration Model }
To begin with,  we give the definition of the  boundary $p(x)$ distribution for a random variable.
\begin{definition}\label{def_bpd}
A   random variable $X \in \Re $ is said to follow a  boundary $p(x)$ distribution  with a PDF $p(x)$ and boundary parameters $\vartheta_l$ and $\vartheta_h$ , if its PDF is given by
\begin{subnumcases}{f_X(x) =}
    p(x)   \ \ \ \ \ \ \ \ \ \ \ \  \ \ \ \ \ \ \    \vartheta_l < x < \vartheta_h, \label{eqn_defpx:a}\\
    \delta(0) \left(\int_{-\infty}^{\vartheta_l}p(x)dx  +  \int^{+\infty}_{\vartheta_h}p(x)dx)\right) \notag\\
     \ \ \ \ \ \ \  \ \ \ \ \ \ \ \   x =\vartheta_l \ \ \text{\rm or} \ \ x=\vartheta_h, \label{eqn_defpx:b} \\
      0   \ \ \ \ \ \ \ \ \ \ \ \  \ \ \ \ \ \ \ \ \    \text{otherwise},  \label{eqn_defpx:c}
  \end{subnumcases}\label{eqn_defpx}
where $|\vartheta_l| \leq 1$, $|\vartheta_h| \leq 1$, and $\delta(\cdot)$ is a Dirac's function.

 \end{definition}
The physical meanings of Definition \ref{def_bpd} are as follows.  When subjected to a random disturbance, the object undergoes a translation or rotation by a random value of $X$ whose PDF is $p(x)$. If the force of the disturbance is sufficiently strong, the object can encounter the boundary, including mechanical obstacles or electronic limits, called boundary events. The second multiplier of~\eqref{eqn_defpx:b} indicates the probability of the boundary event occurring. Since the Dirac's function $\delta(0) = \frac{\int_{-\infty}^\infty \delta(x)dx}{dx}=\frac{1}{dx}$  is often used to study  discrete signals, the continuous random variable $X$  can be treated as a discrete random variable when $X =\vartheta_l$ or $X=\vartheta_h$. Thus, Definition \ref{def_bpd} is an effective means to solve the statistical issues  with boundary effects.

 Suppose the PDF of AVI $X_i=Y^2_i$ within $(\vartheta_l, \vartheta_h)=(0,1)$ is $p(x)$, the AVI $X_i\!=\!Y_i^2 \geq 0$ obeys the  boundary $p(x)$ distribution.  The PDF of $X_i$ is
\begin{subnumcases}{f_{X_i}(x)\!=\! }
     p(x)    & $\ 0 < x < 1$, \label{eqn_xpdf:a}\\
    \delta(0)  ( 1 - \int^{1}_{0}p(x)dx)   & $ x = 1$, \label{eqn_xpdf:b}\\
      0     & $\text{otherwise}$, \label{eqn_xpdf:c}
\end{subnumcases}\label{eqn_xpdf1111}where~\eqref{eqn_xpdf:a} means that $Y_i$ follows the distribution with the PDF $p(x)$ in the range of $(0, 1)$. In  practical engineering, the antenna is often blocked when the angle of the antenna vibration arrives $90^\circ$.  Thus, Eq. \eqref{eqn_xpdf:b} means that the probability of the blocked event occurring is $\int^{\infty}_{1}p(x)dx$. According to the physical meaning of the PDF,  the PDF at $X_i$=$1$ is $\frac{\mathbb{P}[X_i=1]} {dx}$=$\frac{1}{dx}\int^{\infty}_{1}p(x)dx$=$\delta(0) (1-\int^{1}_{0}p(x)dx)$.
From~\eqref{eqn_xpdf:a}--\eqref{eqn_xpdf:c} and Lemmas \ref{prop_jbiconvex}--\ref{prop_jbiXeven}, we can obtain the CDF and PDF of the channel power gain as following theorem.
\begin{theorem}\label{prop_jbicdf}
If the AVI $X_i\!=\!Y_i^2$ follows  boundary $p(x)$ distribution with the boundary 1 as~\eqref{eqn_xpdf:a}--\eqref{eqn_xpdf:c}, the CDF and PDF of the MI fast-fading   gain satisfy
%
%

  ${F_{J_{b,i}}(z)\! = \! }$
 \vspace{-1em}

 \begin{subnumcases}{}
     1 \ \ \ \ \ \ \ \ \ \ \ \ \ \ \ \ \ \ \ \ \ \ \ \ z \geq 1,  \label{eqn_jbicdf:a}\\
    \tfrac{2 \! - \! \int_{X_{i\mathrm{L}}(z)}^{X_{i\mathrm{H}}(z)}p(x)dx}{2} \ \  \max\{ \sin^2\phi_i, \cos^2\phi_i\} \!\leq \! z \!< \!1, \label{eqn_jbicdf:b}\\
     \tfrac{1}{2}\int^{X_{i\mathrm{L}}(z)}_{0}p(x)dx   + \tfrac{1}{2}\int^{X_{i\mathrm{H}}(z)}_{0}p(x)dx \notag\\
                       \ \ \ \ \ \ \ \ \ \ \ \ \ \ \ \ \cos^2\phi_i \!\leq\!  z \!<\!  \sin^2\phi_i, \label{eqn_jbicdf:c}\\
      1\!-\!\tfrac{1}{2}\int^{ X_{i\mathrm{L}}(z)}_{0}p(x)dx \!- \!\tfrac{1}{2}\int^{X_{i\mathrm{H}}(z)}_{0}p(x)dx \notag\\
                    \ \ \ \ \ \ \ \ \  \ \ \ \ \ \ \    \sin^2\phi_i \!\leq\!  z \!<\! \cos^2\phi_i, \label{eqn_jbicdf:d}\\
     \tfrac{\int^{X_{i\mathrm{H}}(z)}_{X_{i\mathrm{L}}(z)}p(x)dx}{2}   
            \ \ \   0 \! \leq  \! z \! <\! \min\{\sin^2\phi_i, \cos^2\phi_i\},   \label{eqn_jbicdf:e}                                                                             \\
      0       \ \ \ \ \ \ \ \ \ \ \ \ \ \ \ \ \ \ \ \ \ \ \ \ z<0, \label{eqn_jbicdf:f}
 \end{subnumcases}\label{eqn_jbicdf11111}
and
  \begin{subnumcases}
  {f_{J_{b,i}}(z)\! = \!}
    \tfrac{| X^{'}_{i\mathrm{L}}(z) | p(X_{i\mathrm{L}}(z))\! +\!   | X^{'}_{i\mathrm{H}}(z)| p(X_{i\mathrm{L}}(z))}{2} \notag\\
                           \ \ \ \ \ \             z \in (0,\sin^2\phi_i) \bigcup (\sin^2\phi_i,1),   \label{eqn_jbipdf:a}\\
    \delta(0)(1 - \int^{1}_{0}p(x)dx)  \ \ \  z = \sin^2\phi_i, \label{eqn_jbipdf:b}\\
      0     \ \ \ \ \ \ \ \ \ \ \ \ \ \ \ \ \ \ \ \ \ \ \ \   \label{eqn_jbipdf:c}\text{otherwise}.
 \end{subnumcases}\label{eqn_jbipdf11111}
respectively. Here  $X_{i\mathrm{L}}(z)$ and $X_{i\mathrm{H}}(z)$ are two solutions of $(J^{\sqrt{.}}_{b,i}(X_i)=0)$ and
\begin{subequations}\label{eqn_JbiInv}
\begin{align}
X_{i\mathrm{L}}(z) &= \cos^2\phi_i- z\cos2\phi_i \!-\! |\sin 2\phi_i|\sqrt{ z \left(1 -z\right)}, \label{eqn_JbiInv:l}\\
X_{i\mathrm{H}}(z) &= \cos^2\phi_i- z\cos2\phi_i \!+\! |\sin 2\phi_i|\sqrt{ z \left(1 -z\right)}, \label{eqn_JbiInv:h}
\end{align}
\end{subequations}
whose derivatives are
\begin{subequations}\label{eqn_JbiInvDeriv}
\begin{align}
X^{'}_{i\mathrm{L}}(z) &= -\cos \! \left(2 \phi_i \right)-\tfrac{{| \sin \! \left(2 \phi_i \right)|} \left(1-2 z \right)}{2 \sqrt{z \left(1-z \right)}}, \\
X^{'}_{i\mathrm{H}}(z) &= -\cos \! \left(2 \phi_i \right)+\tfrac{{| \sin \! \left(2 \phi_i \right)|} \left(1-2 z \right)}{2 \sqrt{z \left(1-z \right)}},
\end{align}
\end{subequations}
respectively.

\end{theorem}

\begin{myproof}
Please refer to Appendix \ref{sect_proofjbicdf}.
\end{myproof}

Besides the PDF and CDF of the  VMI fast-fading gain, the expectation of the VMI fast-fading  gain is also an important parameter in  the study of the mobile MI systems. It may make a remarkably decreasing  or increasing in the transmission rate and communication distance.

Since the PDF of  $J_{b,i}$ is not continuous at  $J_{b,i} = J^{\sqrt{.}}_{b,i}(1)$, we can treat such $J_{b,i}$ as a discrete random variable  whose probability is $\lim\limits_{dz\rightarrow 0}f_{J_{b,i}}(1)dz=\frac{f_{J_{b,i}}(1)}{\delta(0)}$.  The  expectation of $J_{b,i}$ can be written as
\begin{equation}\label{eqn_EJbi0}
\begin{aligned}
\mathbb{E}(J_{b,i})&\!= \!\int_{0}^{1} J_{b,i} f_{J_{b,i}}(z)dz
                   \! +\!  J^{\sqrt{.}}_{b,i}(1) \mathbb{P}[J_{b,i}=J^{\sqrt{.}}_{b,i}(1)]\\
                   &\!=\! \!\int_{0}^{1} J_{b,i} f_{J_{b,i}}(z)dz
                   \! +\!  \sin^2\phi_i (1 \!-\! \int^{1}_{0}p(x)dx).
\end{aligned}
\end{equation}
Unfortunately, due to the inverse functions and their derivatives in~\eqref{eqn_jbipdf:a}--\eqref{eqn_jbipdf:c}, the expression~\eqref{eqn_EJbi0} imply that it may be quite difficult to simplify the integral. Such integral may reduce the convergence speed of related algorithms. We observe that the first item $\int_{0}^{1} J_{b,i} f_{J_{b,i}}(z)dz \triangleq \mathcal{E}_{\rm ct}$  of~\eqref{eqn_EJbi0} can be further deduced into
\begin{equation}\label{eqn_EJbi01}
\begin{aligned}
 \mathcal{E}_{\rm ct} &= \int_{-1}^{0} J^{\sqrt{.}}_{b,i}(y) f_{Y_i}(y)dy +  \int_{0}^{1} J^{\sqrt{.}}_{b,i}(y) f_{Y_i}(y)dy       \\
                    &= \int_{0}^{1} J_{b,i-}(x) \tfrac{f_{X_i}(x)}{2}dx +  \int_{0}^{1} J_{b,i+}(x) \tfrac{f_{X_i}(x)}{2}dx, \\
                    &= \frac{1}{2}\int_{0}^{1} \cos^2\phi_i p(x) - \cos (2\phi_i) x p(x)dx.
\end{aligned}
\end{equation}
 After substituting~\eqref{eqn_xpdf:a}--\eqref{eqn_xpdf:c} and~\eqref{eqn_EJbi01} into~\eqref{eqn_EJbi0},   the expectation of the VMI fast-fading gain is simplified as
\begin{equation}\label{eqn_EJbi}
\begin{aligned}
\mathbb{E}(J_{b,i}) \!=\! \cos 2\phi_i\int_{0}^{1} p(x)dx
                    \!- \!\cos2\phi_i \int_{0}^{1} x p(x)dx \!+\!\sin^2\phi_i             .
\end{aligned}
\end{equation}
We see that $\int_{0}^{1} p(x)dx$ and $\int_{0}^{1} x p(x)dx$  are the forms of the CDF and  expectation of  $X_i$, respectively. Namely, the convergence of~\eqref{eqn_EJbi} are directly determined by the convergence of the CDF and expectation of $X_i$. Therefore, we can achieve better convergence performance from~\eqref{eqn_EJbi} than~\eqref{eqn_EJbi0}.

Theorem \ref{prop_jbicdf}  presents the statistical characteristics of the VMI fast-fading. However, the CDF of VMI fast-fading gain depends on  the statistical characteristics of road roughness and vehicle vibration system. Specifically, the vehicle vibration system  is  complicated. The vibration distribution may be with  quite complex PDFs. We can use the numerical methods to calculate the integral in Theorem \ref{prop_jbicdf} and get the approximate VMI fast-fading gain.


Although studying the vehicle vibration system is beyond the scope of this paper, numerous studies have proposed approximate and simplified models for vehicle vibration systems, including the Gaussian road roughness model and the white noise velocity input\cite{Tuerkay2005A,Luo2023Analytical}. Thus, we consider the scenarios under the Gaussian road roughness model.

\subsection{Gaussian Road Roughness Model}

Some researches such as \cite{Tuerkay2005A}, \cite{Luo2023Analytical} approximate  the vehicle vibration system   as a linear system. The response
 to
 disturbances is modeled as the white noise velocity input.  Thus, the NVVO $Y_i$ follows the  normal distribution ($N(0, \sigma_i^2$)) in the range of $(-1, 1)$. Based on~\eqref{eqn_xpdf:a}--\eqref{eqn_xpdf:c}, the AVI $X_i$=$Y_i^2$ follows the central Chi-square with 1 degree in the range of $(0,1)$. 
 Considering the boundary obstacles and using Definition \ref{def_bpd} by $p(x)=\frac{\exp(-\frac{x}{2\sigma_i^2})}{\sqrt{2\pi\sigma_i^2 x}}$, the AVI $X_i$  follows  so-called \emph{boundary central Chi-square} (BCS) distribution whose PDF is
\begin{subnumcases}
 {f_{X_i}(x) =} \frac{1}{\sqrt{2\pi\sigma_i^2 x}}\exp(-\frac{x}{2\sigma_i^2}) \ \ \ \  x\in [0, 1), \label{eqn_xnormpdf:a} \\
    \delta(0)  \left( {\rm 1- erf}\left(\sqrt{\tfrac{x}{2\sigma_i^2 }}\right)\right) \ \ \ \ \  x=1, \label{eqn_xnormpdf:b}\\
      0      \ \ \ \ \ \ \ \ \ \ \ \ \ \ \ \   \text{otherwise},       \label{eqn_xnormpdf:c}
\end{subnumcases}\label{eqn_xnormpdf}where ${\mathrm{erf}(\cdot)}$ is an error function and the variance $\sigma_i^2$ is called \emph{average AVI}.   Thus, we can get the following theorem.


\begin{theorem}\label{prop_jbicdfnorm}
If the AVI $X_i$ follows BCS distribution with the PDF as~\eqref{eqn_xnormpdf:a}--\eqref{eqn_xnormpdf:c}, the CDF and PDF of the VMI fast-fading  gain satisfy
\begin{subnumcases} { F_{J_{b,i}}(z)\!=\!}
1 \ \ \ \ \ \ \ \ \ \ \ \ \ \ \ \ \ \ \ \ \ \ \ \ z \geq 1,   \label{eqn_jbicdfnorm:a}\\
 1 - \frac{1}{2} \left[ \mathrm{erf}\left(\sqrt{\tfrac{X_{i\mathrm{H}}(z)}{2\sigma_i^2}} \right) - \! \mathrm{erf}\left(\sqrt{\tfrac{X_{i\mathrm{L}}(z)}{2\sigma_i^2}} \right) \right]  \ \notag \\
      \ \ \ \ \ \ \ \ \   \max\{ \sin^2\phi_i, \cos^2\phi_i\} \!\leq \! z \!< \!1,  \label{eqn_jbicdfnorm:b}\\
 \frac{1}{2} \left[ \mathrm{erf}\left(\sqrt{\tfrac{X_{i\mathrm{H}}(z)}{2\sigma_i^2}} \right) + \! \mathrm{erf}\left(\sqrt{\tfrac{X_{i\mathrm{L}}(z)}{2\sigma_i^2}} \right) \right] \notag\\
                       \ \ \ \ \ \ \ \ \ \ \ \ \ \ \ \ \ \ \ \ \cos^2\phi_i \!\leq\!  z \!<\! \sin^2\phi_i, \label{eqn_jbicdfnorm:c} \\
  1\!-\! \frac{1}{2} \left[ \mathrm{erf}\left(\sqrt{\tfrac{X_{i\mathrm{H}}(z)}{2\sigma_i^2}} \right) + \! \mathrm{erf}\left(\sqrt{\tfrac{X_{i\mathrm{L}}(z)}{2\sigma_i^2}} \right) \right]\notag \\
                    \ \ \ \ \ \ \ \ \ \ \ \ \ \ \ \ \ \ \ \   \sin^2\phi_i \!\leq\!  z \!<\! \cos^2\phi_i,  \label{eqn_jbicdfnorm:d} \\
      \frac{1}{2} \left[ \mathrm{erf}\left(\sqrt{\tfrac{X_{i\mathrm{H}}(z)}{2\sigma_i^2}} \right) - \! \mathrm{erf}\left(\sqrt{\tfrac{X_{i\mathrm{L}}(z)}{2\sigma_i^2}} \right) \right]  \notag\\
       \ \ \ \ \ \ \ \     0 \! \leq  \! z \! <\! \min\{\cos^2\phi_i, \sin^2\phi_i\} ,                          \label{eqn_jbicdfnorm:e}                                                      \\
      0       \ \ \ \ \ \ \ \ \ \ \ \ \ \ \ \ \ \ \ \ \ \ \ \ z<0,  \label{eqn_jbicdfnorm:f}
\end{subnumcases}\label{eqn_jbicdfnorm111}
 and
   \begin{subnumcases}{f_{J_{b,i}}(z)\! = \!} \label{eqn_jbipdfnorm111}
   \frac{1}{2} \left(  { \tfrac{| X^{'}_{i\mathrm{L}}(z) | e^{-\tfrac{X_{i\mathrm{L}}(z)}{2\sigma_i^2}}}{\sqrt{2\pi\sigma_i^2 X_{i\mathrm{L}}(z)}}  + \tfrac{| X^{'}_{i\mathrm{H}}(z) |  e^{-\tfrac{X_{i\mathrm{H}}(z)}{2\sigma_i^2}}}{\sqrt{2\pi\sigma_i^2 X_{i\mathrm{H}}(z)}}}\right)  \notag \\
           \ \ \ \ \ \ \ \    \ \ \ \ \ \ \ \  \ \ \ \ \           \ \ \                      z \in (0,1),   \label{eqn_jbipdfnorm:a} \\
    \delta(0)  \left(1 \!-\! \mathrm{erf}\left(\sqrt{\tfrac{1}{2\sigma_i^2}} \right)\right)   \ \ \  z = \sin^2\phi_i, \label{eqn_jbipdfnorm:b}\\
      0      \ \ \ \ \ \ \ \  \ \ \ \ \ \ \ \ \ \ \ \ \   \ \ \ \ \ \ \text{otherwise}.  \label{eqn_jbipdfnorm:c}
 \end{subnumcases}
respectively. Here  $X_{i\mathrm{L}}(z)$ and $X_{i\mathrm{H}}(z)$ are given by~\eqref{eqn_JbiInv}, and their derivatives ($X^{'}_{i\mathrm{L}}(z)$ and $X^{'}_{i\mathrm{H}}(z)$) are given by~\eqref{eqn_JbiInvDeriv}.

\end{theorem}

\begin{myproof}
After substituting $p(x)=\frac{1}{\sqrt{2\pi\sigma_i^2 x}}\exp(-\frac{x}{2\sigma_i^2})$ into Theorem \ref{prop_jbicdf}, Eq. ~\eqref{eqn_jbicdfnorm:a}--\eqref{eqn_jbicdfnorm:f} and~\eqref{eqn_jbipdfnorm:a}--\eqref{eqn_jbipdfnorm:c} hold.

This concludes the proof.
\end{myproof}

Similarly, we substitute $p(x)$$=$$\frac{1}{\sqrt{2\pi\sigma_i^2 x}}\exp(-\frac{x}{2\sigma_i^2})$ into~\eqref{eqn_EJbi} and obtain the expectation of the VMI fast-fading gain
\begin{equation}\label{eqn_EJbiNorm}
\begin{aligned}
\mathbb{E}(J_{b,i}) & \!=\! \left( \cos \! \left(2 \phi_i \right)\!-\! \sigma_i^{2} \cos \! \left(2 \phi_i \right)\right) \mathrm{erf}\! \left(\sqrt{\tfrac{1}{2 \sigma_i^{2}}}\right) \\
 &+ \sqrt{\tfrac{2 \sigma_i^{2} }{\pi}}\, \cos \! \left(2 \phi_i \right){\mathrm e}^{-\frac{1}{2 \sigma_i^{2}}}+ \sin^2\phi_i.
\end{aligned}
\end{equation}

From~\eqref{eqn_EJbiNorm}, an important performance characteristic  can be obtained as below.

\begin{proposition}\label{prop_jbiej2normprop}
If the AVI $X_i$ follows the BCS distribution with the PDF as~\eqref{eqn_jbipdfnorm111}--\eqref{eqn_jbipdfnorm:c}, the expectation of MI fast-fading  gain decreases with the increase of the average AVI $\sigma^2_i$ when the ABO $\phi_i\in[ k\pi,\frac{\pi}{4}+k\pi)$, and increases with the increase of  $\sigma^2_i$  when  $\phi_i\in(\frac{\pi}{4}+k\pi, \frac{\pi}{2}+k\pi]$.
\end{proposition}

\begin{myproof}
Please refer to Appendix \ref{sect_proofjbiej2normprop}.
\end{myproof}

\vspace{0.3em}
 Proposition \ref{prop_jbiej2normprop} indicates that VMI fast-fading with a large ABO has a positive effect on the VMI channel performance. Conversely, the VMI fast-fading gain with a small ABO has a negative effect  on the  VMI channel performance. In  real vehicle systems, as the vibration angle $\theta_i$ seldom  reaches $\frac{\pi}{2}$, the average AVI  $\sigma_i^2$ may not be larger than 1. Therefore, an advantage for  up-layer network protocols can be obtained as follows.
\begin{remark}\label{rmk_ejbiWellDistribute}
In realistic scenarios,  the VMI fast fading can enhance the small channel power gains while reducing the large channel power gains.  That is to say, the VMI fast-fading gain may make the VMI channel coefficients around the base station more uniform. 
\end{remark}

The VMI fast-fading gain may also bring the challenges of the outage probability problems, detailed in below.

\subsection{Outage probability }

Outage probability is defined as the probability that a given information rate $\mathcal(R)$ is not supported\cite{Hong2010Cooperative}. It is equivalent to the probability of the instantaneous receive  signal-to-interference-plus-noise ratio (SINR) being below a threshold $\Upsilon_{\rm th}$, i.e.,
\vspace{-0.0em}
\begin{equation}\label{eqn_pout0}
\begin{aligned}
 \rho_{b,i}(\Upsilon_{\rm th})&\triangleq\!\mathbb{P}\left[\Upsilon_{b,i}\!<\!\Upsilon_{\rm th} \right].
\end{aligned}
\vspace{-0.0em}
  \end{equation}
where $\Upsilon_{b,i}$ is the instantaneous received SINR for link $b\rightarrow i$ and given by
\begin{equation}\label{eqn_sinr}
\begin{aligned}
 \Upsilon_{b,i} = \frac{P_{b,i} \mathcal{G}_{\rm co}\mathcal{M}^2_{b,i}J_{b,i}}{P_{\rm no}+\sum\limits_{\imath\in\Theta}P_{\imath,i} \mathcal{G}_{\rm co}\mathcal{M}^2_{\imath,i}\mathbb{E}(J_{\imath,i})}.
\end{aligned}
\vspace{-0.0em}
  \end{equation}
where $P_{i1,i2}$ denotes the transmit power spectral density (PSD) from node $i1$ to $i2$,  $\Theta$ is the  set of the interference nodes, $P_{\rm no}$ is the noise PSD.  $P_{\imath,i} \mathcal{G}_{\rm co}\mathcal{M}^2_{\imath,i}\mathbb{E}(J_{\imath,i})\triangleq\mathcal{I}_{\imath,i}$ represents the received interference PSD from the interference node $\imath$.
As the interference power $\mathcal{I}_{\imath,i}$ of the  node $i$ is much smaller than the received signals,  $\mathcal{I}_{\imath,i}$ is considered as a deterministic variable.  We use
the expectation $\mathbb{E}(J_{\imath,i})$ in~\eqref{eqn_sinr} instead of the instantaneous  random value $J_{\imath,i}$ to simplify the model of the received SINR.  Accordingly, after solving the inequality $\Upsilon_{b,i}\!<\!\Upsilon_{\rm th}$, we have
\vspace{-0.0em}
\begin{equation}\label{eqn_poutdown}
\hspace{-0.4em}\begin{aligned}
 \rho_{b,i}(\Upsilon_{\rm th}) &\!\simeq  \! \mathbb{P}\left[J_{b,i}\!<\!\tfrac{\Upsilon_{\rm th}(P_{\rm no}+\sum\limits_{\imath\in\Theta}P_{\imath,i} \mathcal{G}_{\rm co}\mathcal{M}^2_{\imath,i}\mathbb{E}(J_{\imath,i})) }{P_{b,i} \mathcal{G}_{\rm co}\mathcal{M}^2_{b,i}} \right]\\
    &= F_{J_{b,i}}\left(\tfrac{\Upsilon_{\rm th}(P_{\rm no}+\sum\limits_{\imath\in\Theta}P_{\imath,i} \mathcal{G}_{\rm co}\mathcal{M}^2_{\imath,i}\mathbb{E}(J_{\imath,i})) }{P_{b,i} \mathcal{G}_{\rm co}\mathcal{M}^2_{b,i}}\right).
  \end{aligned}
\end{equation}

Similar to the EMW based  communication, it can be concluded from~\eqref{eqn_poutdown} that the higher transmitting power can bring benefit to the outage probability performance.  While different from the  EMW communication,  we see an  interesting phenomenon as following. Since $F_{J_{b,i}}(\sigma^2_i)$ is not a monotonic function about $\sigma^2_i$ (see theorem \ref{prop_jbicdfnorm}), the larger average AVI may not necessarily  reduce the  outage probability performance. That is to say, there are  a few opportunities to  improve overall VMI network  performance with  a suitably large average AVI.

\section{ Power Control Optimization Algorithm With Multiagent Q-Learning }\label{sect RL}
As the average AVI is highly unpredictable due to its significant dependence on the driver's mind, the methods solving the  PCO problems for the traditional MICs  without fast-fadings are ineffective in resolving   those  with fast-fadings.  In this section, we propose a  CVMC-PCO with multiagent Q-learning and NE game  for high overall  throughput.

\subsection{ Network Modeling }\label{subsect network_model}

We consider a CVMC network as shown in Fig. \ref{fig_network} where there are  $N$ sub-channels and $K$  cells. Each cell has one CVMC BS and $N$  VUs.  The BS in $k$-th cell (sub-network) is denoted by $\mathrm{ND}_k^{(t)} \in \{\mathrm{ND}_1^{(t)}, \mathrm{ND}_2^{(t)}, ... , \mathrm{ND}_K^{(t)}\}$ where $t$ denotes time slot, $t\geq0$ and $k>0$ are integers. The $n$-th  VU in cell $k$ is denoted by   $\mathrm{ND}_{k,n}^{(t)} \in \{\mathrm{ND}_{k,1}^{(t)}, \mathrm{ND}_{k,2}^{(t)}, ... , \mathrm{ND}_{k,N}^{(t)}\}$. In this paper, we do not care about the geometry of the cell (e.g., hexagonal, oval and rectangle). During time slot 0, we assume that the geographical distribution of the VUs follows the Poisson point process  model in the corresponding cell.   When $t>0$, each  VU  drives with a random walk fashion in its corresponding cell, respectively. In addition, to support the multiuser communications, we use the  protocol of time division multiple access (TDMA). Without loss of generality, we assume that the $n$-th sub-channel is allocated to $n$-th VU.

\begin{figure}[t]
        \centering
        \includegraphics[width=2.9in,height=1.6in]{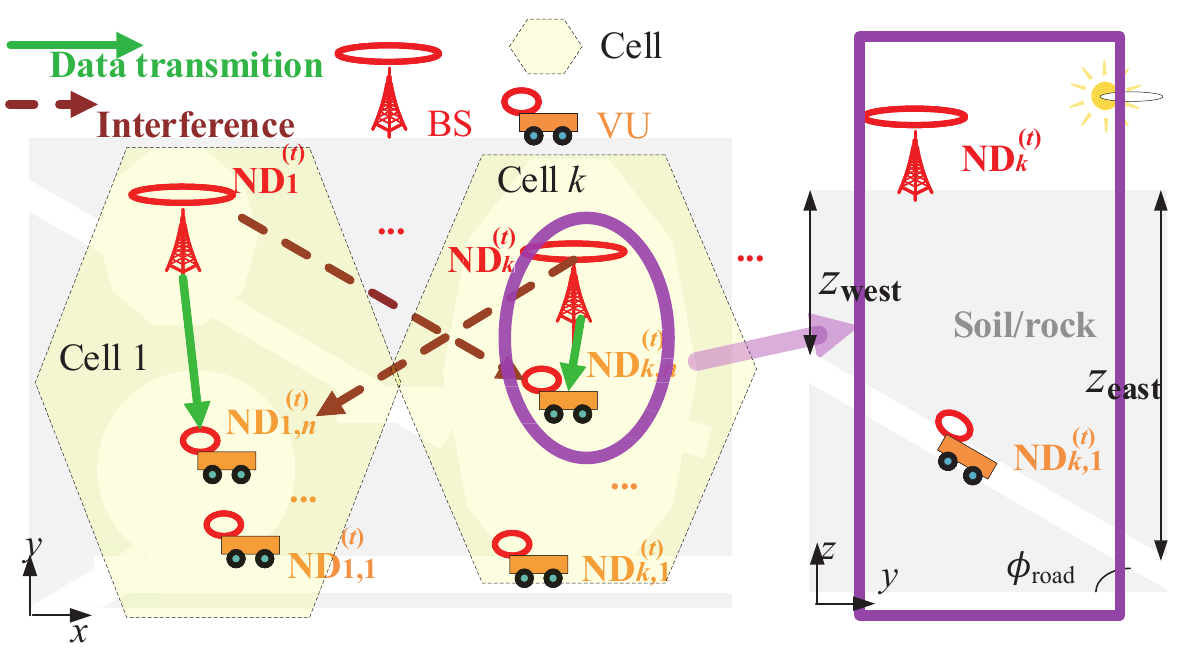}\\
        \vspace{-0.0em}
        \caption{ Example of a CVMC  network.  
 }\label{fig_network}
 \vspace{-1.001em}
    \end{figure}

In this paper, the goal of each cellular sub-network $k \in \mathcal{K}$ is to maximize  the throughput, that is,
\begin{equation}\label{eqn_sbs_goal}
\begin{aligned}
  \hspace{-0.1em}  &\max\limits C_k\!=\!\max\sum\limits_{n=1}^{N}  C_{[k,n]}(1 - \rho_{k,[k,n]}),          \\
     & \ \ s.t. \  \mathbb{I}_n(n_k)\Upsilon_{[k,n]}\!\geq\!\mathbb{I}_n(n_k)\Upsilon_{\mathrm{th}_k},    \\
     &  \ \ \ \ \ \ \ \ \sum\limits_{n}^{N}\mathbb{I}_n(n_k)P_{k} < P^{\rm(sum)}_{k},  \ \  \forall n\!\in\! \mathcal{N}.
     \vspace{-0.1em}
\end{aligned}
\vspace{-0.2em}
\end{equation}
where $P^{\rm(sum)}_{k}$ is the total budget of the PSD for the BS $k$,
$C_{[k,n]}$$=$$\mathbb{E}\{\log_2(1$$+$$\Upsilon_{[k,n]})\}$ represents the spectral efficiency, $\Upsilon_{[k,n]}$ indicates the  average received SINR and
 \begin{equation}\label{eqn_sinrkn}
\begin{aligned}
\Upsilon_{[k,n]} \simeq \frac{P_{k,n} \mathcal{G}_{\rm co}\mathcal{M}^2_{k,[k,n]}\mathbb{E}(J_{k,[k,n]})}{P_{\rm no}+\sum\limits_{\imath\in\mathcal{K} \backslash k}P_{\imath,n} \mathcal{G}_{\rm co}\mathcal{M}^2_{\imath,[k,n]}\mathbb{E}(J_{\imath,[k,n]})}.
\end{aligned}
\vspace{-0.0em}
  \end{equation}
  $P_{k,n}$$=$$\mathbb{I}_n(n_k)P_k$ is the transmitting PSD, $\mathbb{I}_n(n_k)$ indicates whether the $n$-th channel is allocated, namely,
\vspace{-0.3em}
\begin{equation}\label{eqn_alloc}
\begin{aligned}
 & \mathbb{I}_n(n_k)\!=\!\begin{cases}
   \begin{aligned}
     & 1   \!  &n=n_k,   \\ 
     & 0   \!  &n\neq n_k,   \\ 
   \end{aligned}
  \end{cases}
  \end{aligned}\notag
\end{equation}
For the link $\mathrm{ND}_{k_1}$$\rightarrow$$\mathrm{ND}_{k_2,n}$, $\mathcal{M}_{k_1,[k_2,n]}$ and  $\mathbb{E} (J_{k_1,[k_2,n]})$ can be obtained by~\eqref{eqn_mbimax} and~\eqref{eqn_EJbiNorm} (or~\eqref{eqn_EJbi}), respectively.   The outage probability $\rho_{k1,[k2,n]}$ can be obtained by~\eqref{eqn_pout0}.

 \subsection{Problem Transformation}

As the goal of each BS  is the competition to  achieve the maximum network throughput,  the  problem~\eqref{eqn_sbs_goal} becomes  a multi-agent optimization problem.   The $k$-th BS uses the power strategy $p_k$$=$$(n_k, P_k) \in \mathfrak{P}_k $$\triangleq$$\mathcal{N}$$\times$$\mathcal{P}_k$
where $P_k$$\in$$\mathcal{P}_k$$\triangleq$$[P^\mathrm{min}_k, P^\mathrm{ max}_k ]$, $P^\mathrm{min}_k$ and $P^\mathrm{ max}_k$ are the lower  and upper bound of PSD, respectively.

Considering the effect of fast fading, we re-write the received SINR $\Upsilon_{[k,n]}$ in~\eqref{eqn_sbs_goal} as
\vspace{-0.0em}
\begin{equation}\label{eqn_snrk}
\begin{aligned}
\Upsilon_{[k,n]}&(p_k, \mathbf{p}_{-k})   \!=\! \\
&\frac{\mathbb{I}_n(n_k) P_k \mathcal{G}_{co} \mathcal{M}_{k,[k,n]}^2\mathbb{E}(J_{k, [k,n]})}{P_{\mathrm{no}}\!+\!\sum\limits_{i\in \mathcal{K} \backslash\{k\}} \mathbb{I}_n(n_i) P_i   \mathcal{G}_{co} \mathcal{M}_{k,[k,n]}^2 \mathbb{E}(J_{i, [k,n]})},
\end{aligned}
\vspace{-0.0em}
\end{equation}
where $\mathbf{p}_{-k}=(p_1, p_2, ..., p_{k-1}, p_{k+1}, ..., p_K) \in \mathfrak{P}_{-k} \triangleq \mathfrak{P}_1\times\mathfrak{P}_2\times ...\times \mathfrak{P}_{k-1}\times \mathfrak{P}_{k+1}\times ... \times \mathfrak{P}_K$. 
Therefore, we have
\vspace{-0.0em}
\begin{equation}\label{eqn_cin}
\begin{aligned}
  \hspace{-0.4em}C_{[k,n]}(p_k, \mathbf{p}_{-k})\!=\!\frac{1\!-\!\rho_{k,[k,n]}}{\daleth}\log_2(1 \! +\! \Upsilon_{[k,n]}(p_k,\mathbf{p}_{-k})),
\end{aligned}
\vspace{-0.0em}
\end{equation}
where $\daleth$ is to compensate the data rate loss caused by the upper layer protocols, and $\mathbf{p}=(p_k, \mathbf{p}_{-k})$.
As  $\mathbf{p}_{-k}$ is presented in the denominator of $\Upsilon_{[k,n]}$, the problem~\eqref{eqn_sbs_goal} is a competitive problem over the cells $k$ and $-k$.
Thus, we define  the instantaneous utility function of cell $k$ as
\vspace{-0.7em}
\begin{equation} \label{eqn_uk}
\begin{aligned}
\hspace{-0.5em}U_k (p_k, \mathbf{p}_{-k}) &\!= \!\sum\limits_{n=1}^{N} (C_{[k,n]}(p_k, \mathbf{p}_{-k})).   \\
\end{aligned}
\vspace{-0.6em}
\end{equation}

In \cite{lin2015distributed}, researchers use the NE strategy to obtain the closed-form expressions of the power strategies for the multi-agent optimization problem over the MI users. The NE is defined as follows.
\vspace{-0.8em}
\begin{definition}\label{def_NE}
The  power strategy vector $(p^*_k, p^*_{-k})$ is said to constitute an NE for the power control game $\mathcal{G}_\mathcal{K}=\{\mathcal{K}, \mathfrak{P}_k, U_k\}$,if for every player $k\in \mathcal{K}$,
\vspace{-0.6em}
\begin{equation}\label{eqn_NE}
U_k(p_k, \mathbf{p}^*_{-k}) \leq U_k(p^*_k, \mathbf{p}^*_{-k}) \ \ \forall p_k \in \mathfrak{P}_k.\notag
\end{equation}
\vspace{-0.5em}
\end{definition}
\vspace{-1.8em}

However, since each VU moves in a random walk fashion with unpredictable $\mathbb{E}(J_{k, [k,n]})$  caused by the unpredictable   vehicle velocity, the instantaneous utility value of each BS greatly varies over tens of seconds,  it is hard to obtain the constant NE strategy through the instantaneous utility value.

%
%
Similar to\cite{Chen2013stochastic}, we  use the  multiagent Q-learning method with the expectation of utility
\vspace{-0.7em}
\begin{equation}\label{eqn_euk}
\begin{aligned}
\widetilde{u}&_k(s_k,a_k, \mathbf{a}_{-k})\!\triangleq  \!\mathbb{E}[\sum\limits_{t=t_0}^\infty \gamma_k^{(t)}  U_k^{(t)}(s_k^{(t)}, p_k, \mathbf{p}^{(t)}_{-k})|s_k^{(t_0)}\!=\!s_k]   \\
 &=\!U_k(s_k, \mathbf{p})\!\sum\limits_{s'_k\in S_k}\mathbb{T}_{s_k,s'_k}\! +\!\gamma_k\!\sum\limits_{s'_k\in S_k}\mathbb{T}_{s_k,s'_k}\widetilde{u}_k (s'_k,a_k, \mathbf{a}_{-k}),
\end{aligned}\notag
\vspace{-0.0em}
\end{equation}
where $s_k^{(t)}$ is  environment state of the $k$-th BS  at time $t$, $a_k = \mathcal{F}_k(p_k)$ is  the action performed by $k$-th BS,  $\mathcal{F}_k$ is a  linear function of power strategy $p_k$,  $\gamma^{(t)}_k$  is the discount factor. $\mathbb{T}_{s_i s'_{i}}$ is the state transition probability  from state $s_k$ to state $s'_{k}$.

Furthermore, since we use the set with 2-D power strategy  much larger than the 1-D strategy set as the one in \cite{Chen2013stochastic}, the multiagent Q-learning with mixed strategies as the one in\cite{Chen2013stochastic} has a dramatically large  action searching space. We use the pure strategies to achieve a faster convergence speed.  The optimization problem~\eqref{eqn_sbs_goal} can be transformed into 
\vspace{-0.0em}
\begin{subequations}\label{eqn_sbs_prb}
\begin{align}
    & \arg\max\limits_{p_k \in \mathfrak{P}_k} \widetilde{u}_k(s_k,a_k, \mathbf{a}_{-k})  \label{eqn_sbs_prb:u0}  \\
   & s.t. \ \ \ \mathbb{I}_n(n_k)\Upsilon_{[k,n]}(a_k, \mathbf{a}_{-k})\!\geq\!\mathbb{I}_n(n_k)\Upsilon_{\rm th}
   \ \  \forall n\in \mathcal{N}.    \label{eqn_psb_prb:st}
\end{align}
\vspace{-0.0em}
\end{subequations}
It has the following proposition.

\vspace{-0.0em}
\begin{proposition}\label{prop_NERL}
  Given the  utility $\widetilde{u}_k$$=$$\widetilde{u}_k(s_k,a_k, \mathbf{a}_{-k})$ for  player $k\in \mathcal{K}$, there exists an NE in the  Q-learning based power control game $\mathcal{G}_\mathcal{\widetilde{K}}=\{\mathcal{K},\mathcal{F}_k(\mathfrak{P}_k),\widetilde{u}_k\}$  with 
  an NE power strategy profile  $\mathbf{a^*}$$=$$(a^*_k, \mathbf{a^*}_{-k})$$\in$$\mathcal{F}_{k}(\mathfrak{P}_{k})\times \mathcal{F}_{-k}(\mathfrak{P}_{-k})$ where $a^*_{k_1} = \mathcal{F}_{k_1}(p^*_{k_1})$, $\mathcal{F}_{(k_1)}$ is a  linear map, and $k_1 \in \mathcal{K}$. 
\end{proposition}

\vspace{-0.0em}
\begin{myproof}
The  sets $\mathcal{P}_k$ and $\mathfrak{P}_k$ are the nonempty, convex and compact subset of Euclidean space. According to~\eqref{eqn_poutdown} and Theorem \ref{prop_jbicdfnorm}, the outage probability $\rho_{k,[k,n]}(P_k)$ is a monotonic decreasing function. According to~\eqref{eqn_snrk}, $\Upsilon_{[k,n]}(P_k)$ and $U_k(P_k)$ are  monotonic increasing functions. Since $\widetilde{u}_k(\cdot)$ and $\mathcal{F}_k(p_k)$ are the linear functions of  $U_k^{(t)}(\cdot)$ and  $p_k$, respectively,  $\widetilde{u}_k(\cdot)$ is  quasi-concave. 

This concludes the proof.
\end{myproof}
\vspace{-0.2em}

\subsection{Design Multiagent Q-Learning  Algorithm}\label{subsect RL}

%
%
%
%
%

According to the Q-learning method\cite{Sutton2018Reinforcement},  the power control problem~\eqref{eqn_sbs_prb} can be solved  through  $Q_k^* (s_k, a_k)$$\triangleq$$ \mathbb{E}\{\widetilde{u}_k(s_k,a^*_k, \mathbf{a}^*_{-k})|a_k^{(t0)}$$=$$a_k^*\}$ which equivalent to
\vspace{-0.0em}
\begin{equation}\label{eqn_Qkt1}
\begin{aligned}
\hspace{-0.6em}Q_k^{(t\!+\!1)}(s_k, a_k) &\!:=\!Q_k^{(t)}(s_k, a_k)\!+\!\alpha^{(t)}\big[U_k(s_k,  a_k, \mathbf{a}_{-k})   \\
     &  +\!\gamma_k \max\limits_{a'_k\in\mathcal{A}_k}Q^{(t)}_k(s'_k, a'_k)\!-\!Q_k^{(t)}(s_k, a_k )\big],
  \end{aligned}
  \vspace{-0.0em}
\end{equation}
where $\alpha^{(t)}$ denotes the learning rate.

 To design an algorithm for~\eqref{eqn_Qkt1},  we need to define the agent, state, action and reward as follows.

1) AGENT: The $k$-th BS is the  $k$-th agent.

2) STATE: The state means the interaction with the communication environment. We define the state for the $k$-th agent as a tuple $s_k=( n^{\rm co}_k, P_k) \in \mathcal{S}_k$ where $n^{\rm co}_k\in \mathcal{N}$ is the number of connecting VU, i.e.,
 \vspace{-0.0em}
  \begin{equation}\label{eqn_nco}
 n_k^{\rm co}\!=\!\sum\limits_{n=0}^N\mathbb{I}_{\rm True}\left(\mathbb{I}_n(n_k)\Upsilon_{k,n}(a_k, \mathbf{a}_{-k}) > \mathbb{I}_n(n_k)\Upsilon_{\rm th}\right).
 \notag\vspace{-0.0em}
 \end{equation}
 Thus, the state set of the agent $k$ is $\mathcal{S}_k = \mathcal{N}\times\mathcal{P}_k$.

3) ACTION:  As BS $k$ would choose the channel $n_k$ and  the PSD level $P_k$, we  discretize the PSD profile $\mathcal{P}_k$ as $a^{\rm po}_k (P_k)$$=$$\left\lfloor\frac{P_k - P_k^{\rm min}}{P_k^{\rm max}- P_k^{\rm min}}\right\rfloor$$W_k$, i.e., $a^{\rm po}_k$$\in$$\mathcal{A}^{\rm po}_k$$=$$\{0,1, ..., W_k\}$.
Thus, we have $\mathcal{F}_k(p_k)$$=$$\mathcal{F}_k(n_k, P_k)$$\triangleq$$(n_k, a^{\rm po}_k (P_k))$.
We define $a_k$$\in$$\mathcal{A}_k$$=$$\mathcal{N}$$\times$$\mathcal{A}^{\rm po}_k$ as the action that agent $k$ would perform. During every iteration  $t$, the Q-learning algorithm  chooses the action $a_k^{(t)}$ for agent $k$ by
\vspace{-0.0em}
 \begin{equation}\label{eqn_choose_at}
  a_k^{(t)} = \arg\max\limits_{a_k \in \mathcal{A}_k} Q^{t}(s_k, a_k)
  \vspace{-0.0em}
  \end{equation}
with the probability $\mathbb{P}_k >0.9$, $\mathbb{P}_k$ is to avoid convergence earlier and being in local optimized solutions.

5) REWARD: The reward of agent $k$ in state $s_k$ is the immediate return when the $k$-th BS  performs $a_k$ and  the others perform $\mathbf{a}_{-k}$. We define the  instantaneous utility $U_k$ as its reward.

Subsequently, we propose the CVMC-PCO algorithm based on the multiagent Q-learning and NE game as shown in Algorithm \ref{alg_qlearning}.  Its convergence is discussed below.

\begin{algorithm}[h]\label{alg_qlearning}
 \KwIn{The BS index $k$}
    Initiate  $Q^{(0)}_k(s_k, a_k), \forall s_k \in S_k, \forall a_k \in \mathcal{A}_k$ \\
  \tcc{\sf Begin learning}
\For{$t:=0$ to $MaximalIterations$ }
{
           Detect the network basic parameters (positions, velocities, $\sigma^{(t)}_{k,n}$ of VUs); \\
           Choose and perform action $a^{(t)}_k$ according to~\eqref{eqn_choose_at} with the probability $\mathbb{P}_{k}$; \\
            Measure the received SINRs $\Upsilon_{k,n}, \forall n\in \mathcal{N}$, with the feedback information of VUs, and calculate $s^{(t)}_k:=( n^{\rm co}_k, P_k)$ according to~\eqref{eqn_nco} and $a^{(t)}_k$; \\
            Calculate the instantaneous  reward $U_k$ according to~\eqref{eqn_uk}; \\
            Update $Q^{(t)}_k(s_k, a_k)$ -value based on~\eqref{eqn_Qkt1};

}
\tcc{\sf End Learning}
\Return the optimal power strategy $a_k$;
\caption{ CVMC-PCO Algorithm. }
\end{algorithm}
\vspace{-1.0em}

\begin{proposition}\label{ther_conv}
Regardless of any initiated  power strategy profile $\mathbf{a}^{(0)}=(a^{(0)}_k, \mathbf{a}^{(0)}_{-k})$ chosen for  $Q_k^{(0)}(s_k, a_k)$,  and $\mathbb{P}_k > 0.9$, Algorithm \ref{alg_qlearning} converges, for $\forall k \in \mathcal{K}$,  in the $K$-players  game $\mathcal{G}_\mathcal{\widetilde{K}}$.
\end{proposition}
\vspace{-0.7em}
\begin{myproof}
According to Algorithm \ref{alg_qlearning}, the BS $k$ chooses action $a_k$ in state $s_k$ at time step $t$ with probability
\begin{equation}\label{eqn_pik}
\begin{aligned}
 & \pi_k(s_k, a_k)\!=\!\begin{cases}
   \begin{aligned}
     & \mathbb{P}_k   \!  &a_{\rm selected}=a_k,   \\ 
     & \frac{1 - \mathbb{P}_k }{|\mathcal{A}_k|}   \!  &a_{\rm selected}\neq a_k.    \\ 
   \end{aligned}
  \end{cases}
  \end{aligned}
\end{equation}

 For $C_{k,n}(p_k, \mathbf{p}_{-k})$, with  $\mathbb{E}(J_{k_1,[k_2, n_2]})$ treated as deterministic variables  over a sufficiently short time (e.g., 1--10 seconds), $h^{\rm VMI}_{k_1k_2}$=$\mathcal{G}_{co} \mathcal{M}_{k_1,[k_2,n_2]}^2 \mathbb{E}(J_{k_1, [k_2,n_2]})$ becomes a special case of channel power gain ($h_{ij}$) in \cite{Chen2013stochastic}.

As $\rho_{k,[k,n]}(P_k)$ is a decreasing function,  we have
\begin{equation}\label{eqn_ruoeq}
\begin{aligned}
0 \leq 1-\rho_{k,[k,n]}(P_k)\leq 1-\rho_{k,[k,n]}(P^*_k) \leq 1
  \end{aligned}
\end{equation}
where $P^*_k$ is the NE solution and 1-$\rho_{k,[k,n]}(P^*_k)$ can be treated as a constant coefficient.
If we  substitute~\eqref{eqn_pik},~\eqref{eqn_ruoeq} and $h^{\rm VMI}_{k_1k_2}$ into Theorem in \cite{Chen2013stochastic}, the contraction mapping operator and fixed point of  Theorem in \cite{Chen2013stochastic} remain.  Algorithm \ref{alg_qlearning} is  convergent.

This concludes the proof.
\end{myproof}

{\color{red}
\begin{figure}[t]
        \centering
        \includegraphics[width=2.8in,height=1.5in]{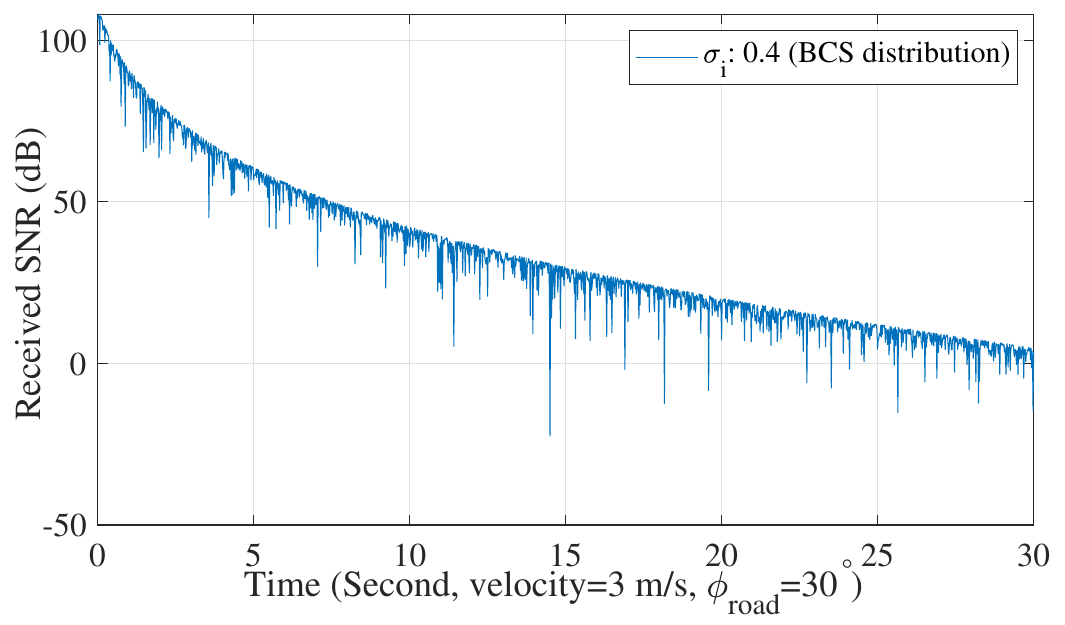}\\
        \vspace{-0.4em}
        \caption{Received SNR under a VMI fast-fading channel. Suppose the vehicle is traveling away from the BS at a speed of 3 m/s along an underground road inclined at a $30^\circ$ angle.
 }\label{fig_snrtime}
 \vspace{-1.001em}
    \end{figure}
    }

\section{Numerical Evaluation }\label{sect sim}

In this section, we validate the CDF, expectation and outage probability of the VMI channel by the Monte Carlo simulations, followed by the  simulations of the convergence performance for  the CVMC-PCO algorithm.

Although our MIC devices, equipped with coils of a 4-meter radius and a transmit power of 40W, can achieve a depth of 300 meters in a coal mine environment, they are not suitable for installation on vehicles. Thus, we modify the parameters of MIC devices for numerical evaluations. Most of these parameters are based on those used in our previous work \cite{Ma2019Antenna} where the MI antennas are used to achieve tens of meters propagation. Namely, we assume  $N_{\mathrm{c}i}$$=$$30$, $a_{\mathrm{c}i}$$=$$0.4m$, $N_{\mathrm{c}b}$$=$$15$, $a_{\mathrm{c}b}$$=$$0.6m$,  $\rho_w$$=$$ 0.0166\Omega/$m, $\mu$$=4\pi$$\times$$10^{-7}$H.m$^{-1}$, $R_L$$=$$R_{\mathrm{c}i}$=$2\pi N_{\mathrm{c}i}a_{\mathrm{c}i}\rho_w$ and $f_0$$=$$10$KHz.

For all evaluations about VMI fast-fading gain, we assume that the vehicle antenna vibration system is based on  the Gaussian road roughness model and white noise velocity input. Namely, the antenna vibration follows the BCS distribution. Fig. \ref{fig_snrtime} illustrates an example of such VMIC fast-fading channel, clearly demonstrating the rapid fluctuations in the instantaneous received SNR.

\subsection{CDF of VMI fast-fading gain}

We validate the CDF of VMI fast-fading gain $J_{b,i}$ by  Monte Carlo's simulations. Firstly, we simulate the ratio  $\frac{\mathrm{counts}(J^\diamond_{b,i}(Y_i)<z)}{5000}$ with 5000 samples of NVVO $Y_{i}$ for each $z\in (0,1]$. In Fig. \ref{fig_cdf}, the dotted lines denote these simulation (i.e., Monte Carlo's simulation) results. Secondly, we compare these results of the ratio  $\frac{\mathrm{counts}(J^\diamond_{b,i}(Y_i)<z)}{5000}$  with the corresponding calculations of~\eqref{eqn_jbicdfnorm:a}--\eqref{eqn_jbicdfnorm:f}. In Fig. \ref{fig_cdf}, the solid lines denote these theoretical calculations. We find that all simulation curves  fit the calculation curves very well, which validates the CDF of the VMI fast-fading gain.

Moreover, the discontinuous points of the curves in Fig. \ref{fig_cdf:1} and Fig. \ref{fig_cdf:2} have the same $z$ value, respectively. Here, the curve C1 ($\sigma_i$$=0$) can be treated as a special case very close to a continuous point. These discontinuous points are brought by the Dirac's function $\delta(0)$ when $z = \sin^2\phi_i$ (see~\eqref{eqn_jbipdfnorm:a}-~\eqref{eqn_jbipdfnorm:c}).

Also, two special curves C1 and C2 in Fig. \ref{fig_cdf} indicate that  $J_{b,i}$ under zero and infinite average AVIs become determined  binary values. Namely, the VMI channel without the antenna vibration remains quasi-static.

\begin{figure}[t]
  \centering
  \subfigure[ $J^{\sqrt{.}}_{b,i}(0) < J^{\sqrt{.}}_{b,i}(1)$.]{
    \label{fig_cdf:1} 
    \includegraphics[width=1.65in,height=1.3in]{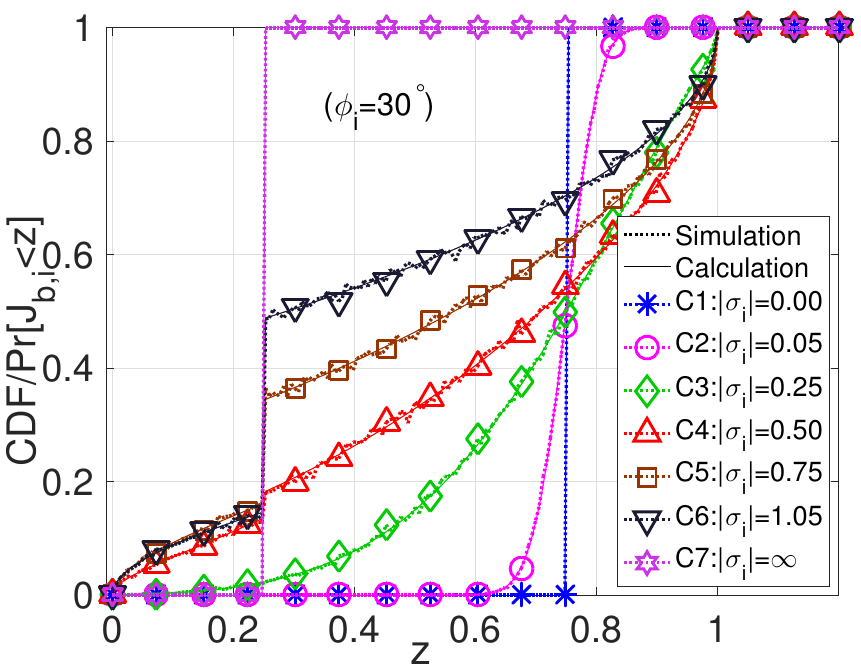}}
  \subfigure[ $J^{\sqrt{.}}_{b,i}(0) > J^{\sqrt{.}}_{b,i}(1)$.]{
    \label{fig_cdf:2} 
    \includegraphics[width=1.65in,height=1.3in]{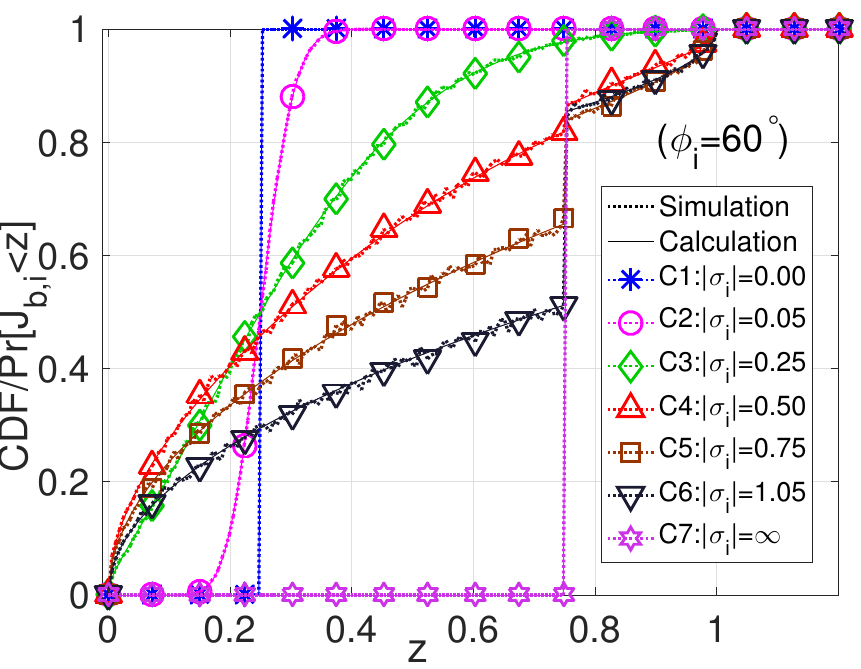}}
  \caption{CDF values of VMI fast-fading gain with different average AVIs. }
  \label{fig_cdf} 
   \vspace{-0.801em}
\end{figure}

\begin{figure}[t]
        \centering
        \includegraphics[width=3.2in,height=2.1in]{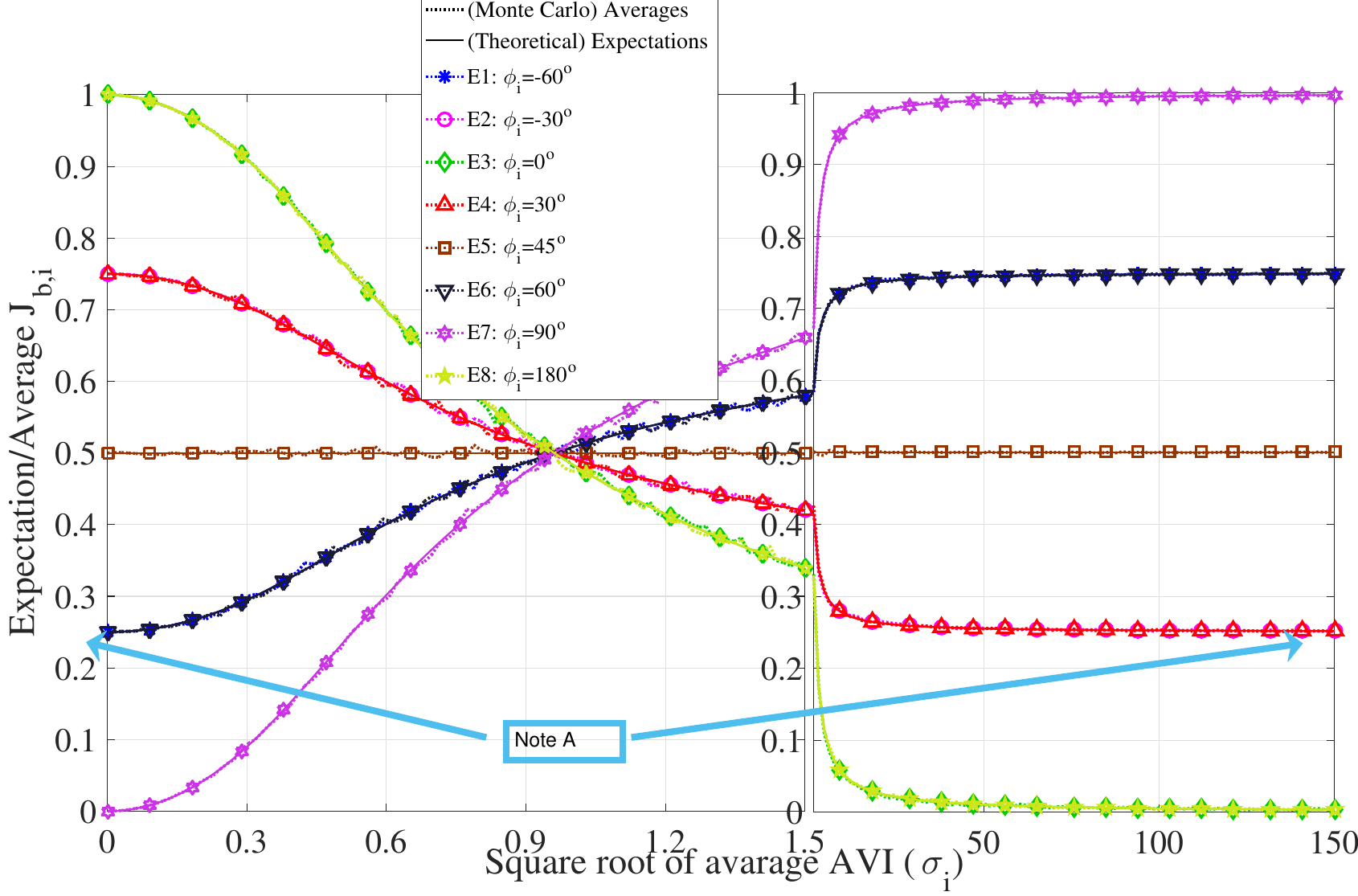}\\
        \vspace{-0.4em}
        \caption{Expectations of the VMI fast-fading with different ABOs $\phi_i$ (Note A: $\mathbb{E}(J_{b,i}|\phi_i)$ is close to $\mathbb{E}(J_{b,i}|\frac{\pi}{2}-\phi_i)$).  
 }\label{fig_ej2howlow}
 \vspace{-1.001em}
    \end{figure}

\vspace{-0.0em}

\subsection{Expectation of VMI fast-fading gain}
  Similar to the validations of VMI fast-fading gain,   we validate the expectation  as shown in Fig. \ref{fig_ej2howlow}. We also find that the theoretical derivations of  the expectations~\eqref{eqn_EJbi} and~\eqref{eqn_EJbiNorm} are valid since all the Monte Carlo's simulation curves are very close
to the corresponding calculation curves.

More importantly, the simulations of $\mathbb{E}(J_{b,i})$ show that the antenna vibration would have a remarkable effect on the average path loss of the VMIC. For instance, from curve E1 in Fig. \ref{fig_ej2howlow}, we observe that the VMI fast-fading gain without antenna vibration ($\sigma_i^2=0$) is $100\%$. When the average AVI increases to 0.95$^2$ (i.e., the average antenna vibration angle is  about $43.5^\circ$ ), the VMI fast-fading gain  decreases to 50\%. While from curve E8 in Fig. \ref{fig_ej2howlow},  the VMI fast-fading gain increases from 0 to 50\% when the average AVI increases from 0 to 0.95$^2$. This phenomenon validates Proposition \ref{prop_jbiej2normprop}.

\begin{figure}[t]
  \centering
  \subfigure[Horizontal distribution.]{
    \label{fig_Ejyz:1} 
    \includegraphics[width=1.60in,height=1.5in]{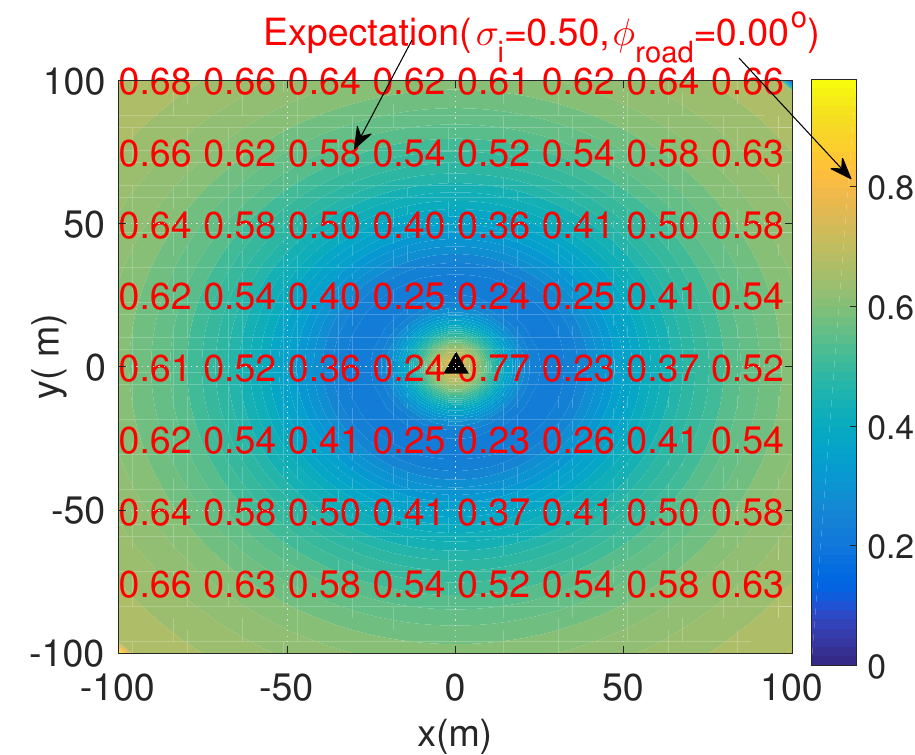}}
  \subfigure[Vertical distribution ($\sigma_i$=$0.5$).]{
    \label{fig_Ejyz:2} 
    \includegraphics[width=1.60in,height=1.5in]{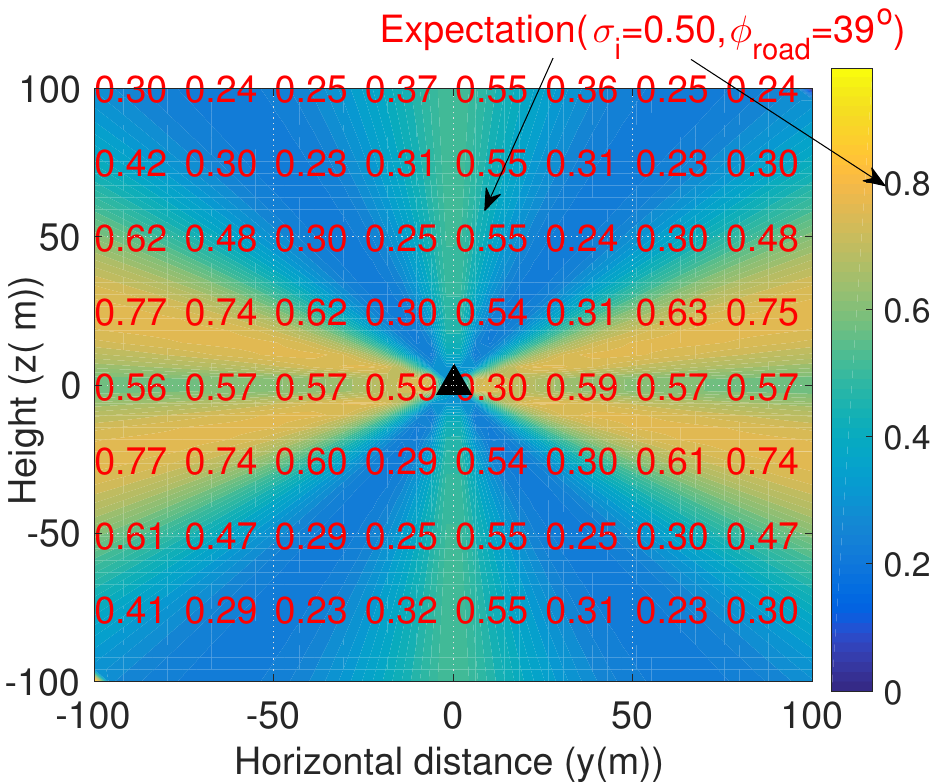}}
    \hspace{1in}
    \subfigure[Vertical distribution ($\sigma_i$=$0.95$).]{
    \label{fig_Ejyz:3} 
    \includegraphics[width=1.60in,height=1.5in]{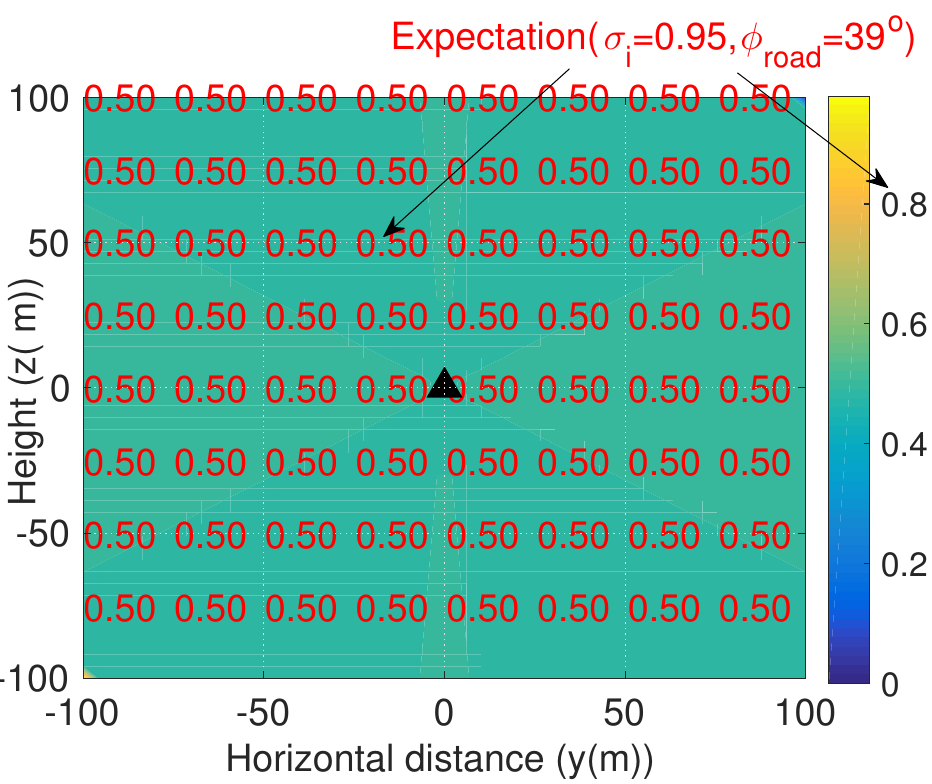}}
      \subfigure[No vibration.]{
    \label{fig_Ejyz:4} 
    \includegraphics[width=1.60in,height=1.5in]{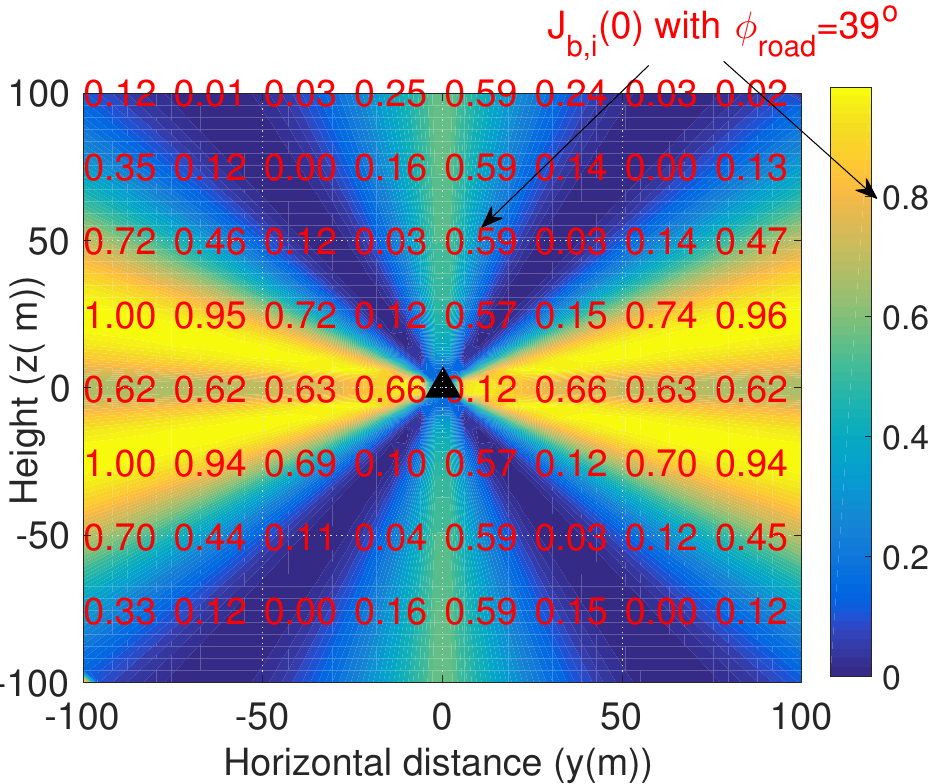}}
  \caption{The spatial distributions of the expectations of the VMI fast-fading gains around the BS, the black triangle denotes the BS. }
  \label{fig_Ejyz} 
  \vspace{-0.9em}
\end{figure}

In actual engineering practice,  the average AVI $\sigma^2_i$ is unlikely to be larger than $0.95^2$. Hence, we further simulate the  spatial distribution of the expectation of the VMI fast-fading gain around the base station with $\sigma^2\leq0.95^2$ (see  Fig. \ref{fig_Ejyz}). Here, Fig. \ref{fig_Ejyz:1} shows that the contour lines of the expectation are the circular shapes, namely, the  horizontal distribution of $\mathbb{E}(J_{b,i})$  depends only on the distance between the BS and VU.  More importantly, Fig. \ref{fig_Ejyz:2}, Fig. \ref{fig_Ejyz:3} and Fig. \ref{fig_Ejyz:4} reveal that the vertical distribution of the  VMI fast-fading gain with the remarkable  antenna vibration is more uniformly distributed than that without antenna vibration,  This phenomenon aligns with our Remark \ref{rmk_ejbiWellDistribute}. The uniformly distributed channel coefficients are beneficial for communication applications but detrimental for localization applications.

\subsection{Outage Probability}

Due to the existence of VMI fast-fading gain, the outage probability of the VMIC cannot be ignored. Fig. \ref{fig_Poutsigma} and  Fig. \ref{fig_Poutpower} show that the remarkable outage probability persists even for the links with very strong signals.

As is shown in Fig. \ref{fig_Poutsigma}, for the most part, the larger average AVIs ($\sigma_i^2$) have more negative effects on the  outage probability performance. However, these simulation results also show the  presence of positive effects from vehicle vibration  when $\sigma_i^2 < 0.4^2$. The reason is that $F_{J_{b,i}}(\sigma_i)$ is not a monotonic function with respect to $\sigma_i$ (see theorem \ref{prop_jbicdfnorm}). In practical  scenarios, the average AVI is often smaller than 0.5 (i.e., average angle $\overline{\theta}_i<45^\circ$). Hence, unlike the  EMW based communications,  a suitably larger average AVI   might  enhance overall VMI network  performance.

 Fig. \ref{fig_Poutpower} shows the effect of BS transmit power on the outage probability. Similar to the EMW based communications, the strong signal can reduce the outage probabilities. In addition, the lines in Fig. \ref{fig_Poutpower} is  discontinuous due to the boundary events of the antenna.

\begin{figure}[t]
        \centering
        \includegraphics[width=3.4in]{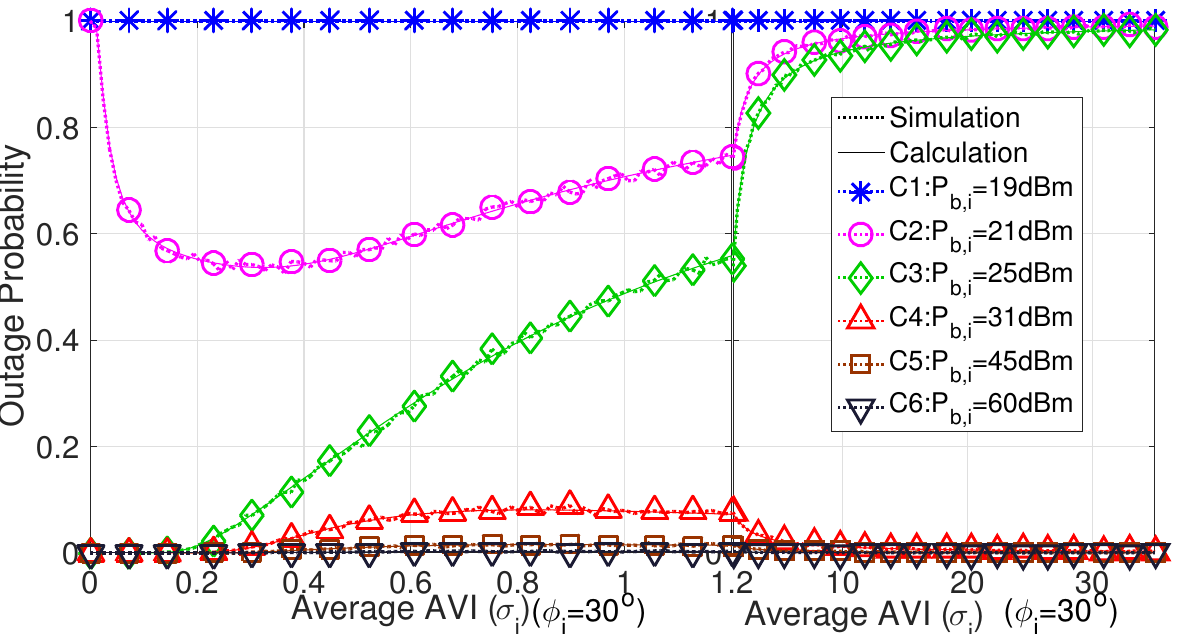}\\
        \vspace{-0.0em}
        \caption{Effects of the average AVI on  outage probabilities under various  transmit powers.  
 }\label{fig_Poutsigma}
 \vspace{-1.0em}
    \end{figure}

\begin{figure}[t]
        \centering
        \includegraphics[width=3.2in]{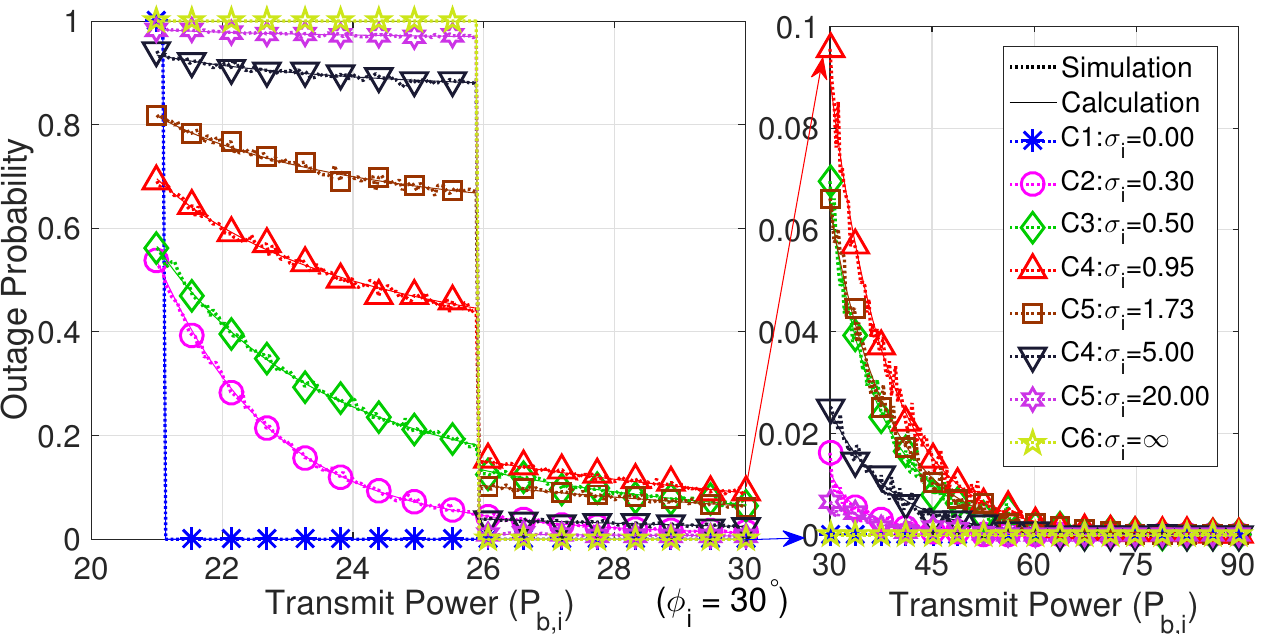}\\
        \vspace{-0.4em}
        \caption{ Effects of the transmit powers on  outage probabilities under various   average AVIs.  
 }\label{fig_Poutpower}
 \vspace{-0.8em}
    \end{figure}

\subsection{Power Control Optimization Algorithm}

\begin{figure}[t]
        \centering
        \includegraphics[width=2.2in,height=1.5in]{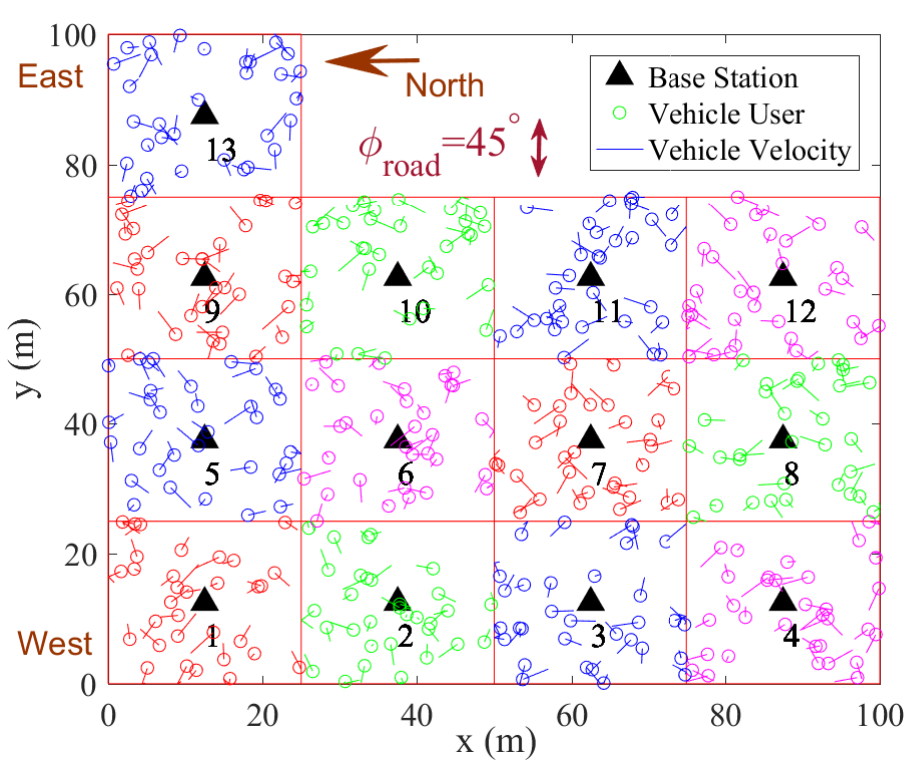}\\
        \vspace{-0.4em}
        \caption{ A CVMC network for CVMC-PCO simulations where $z_{\rm west}$=$10$m (see Fig. \ref{fig_network}), $\phi_{\rm road}$$\simeq$$45^\circ$,  $K$$=$$13$ and $N$$=$$32$. Each VU  moves in a random walk with the maximal velocity of $2$m/s. Its average AVI $\sigma^2_{[k,n]}$ varies randomly at the end of every time $25t$.  
 }\label{fig_pppv32}
 \vspace{-0.9em}
    \end{figure}
For Algorithm 1, we set $\gamma_k$$=$$0.9$, $\alpha_t$$=$$\frac{0.001}{1.000001^{t}}$, $\mathbb{P}_{k}$=$0.9$ and $\mathcal{A}^{\rm po}_k=\{6.2,7.5,8.7,9.9,11.1\}$W. The CVMC network is as depicted in Fig. \ref{fig_network} and Fig. \ref{fig_pppv32} where $K$=$13$ cells  are deployed within a west-east tilting underground space of $100$m$\times$$100$m. This underground space has the depths of $z_{\rm west}$=$-10$m and $z_{\rm east}$=$-110$m (i.e., $\phi_{\rm road}$=$45^\circ$). To simplify matters, we assume that the shape of each MI cell is a square shape instead of a hexagonal one. Each cell is of identical size, containing an MI BS at its center. $N$=$32$ VUs  move in a random walk with a maximal velocity of $2$m/s. The spatial distribution of these VUs follows a homogeneous Poisson Point Process. For each VU, the average AVI $\sigma^2_{[k,n]}$ is unpredictable and varies randomly at the end of every time $25t$.
Additionally, there are $N$=32 TDMA sub-channels for VMICs.  We simulate the case of $N$=1 which can be treated as a fair polling scheme. Namely, every VU has a time-slot of equal duration within a TDMA superframe.

\begin{figure}[t]
  \centering
  \subfigure[$N$=32 ]{
    \label{fig_erw:32} 
    \includegraphics[width=1.60in,height=1.5in]{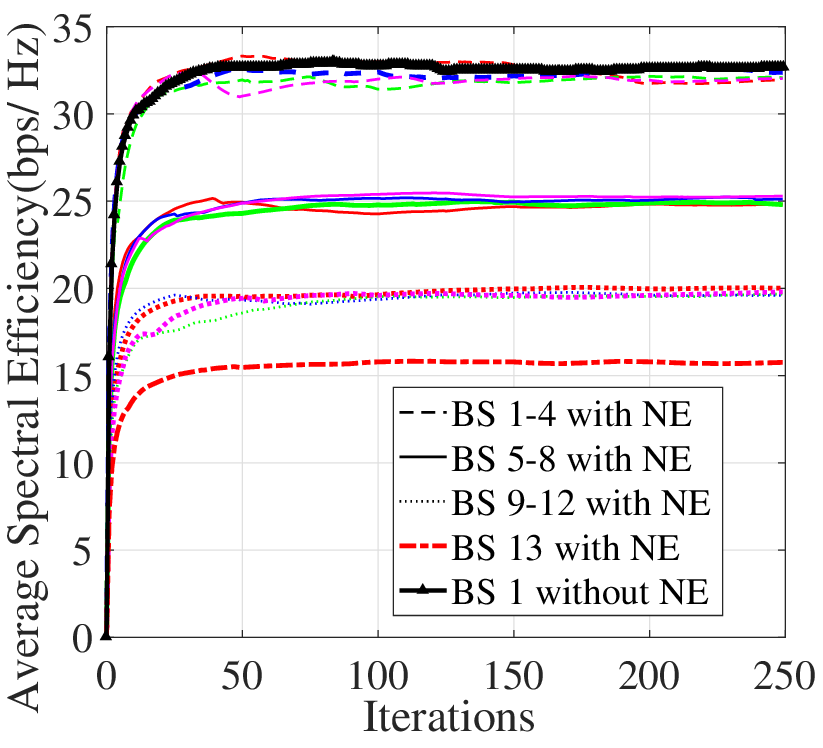}}
  \subfigure[$N$=1 (fair polling scheme)]{
    \label{fig_erw:1} 
    \includegraphics[width=1.60in,height=1.5in]{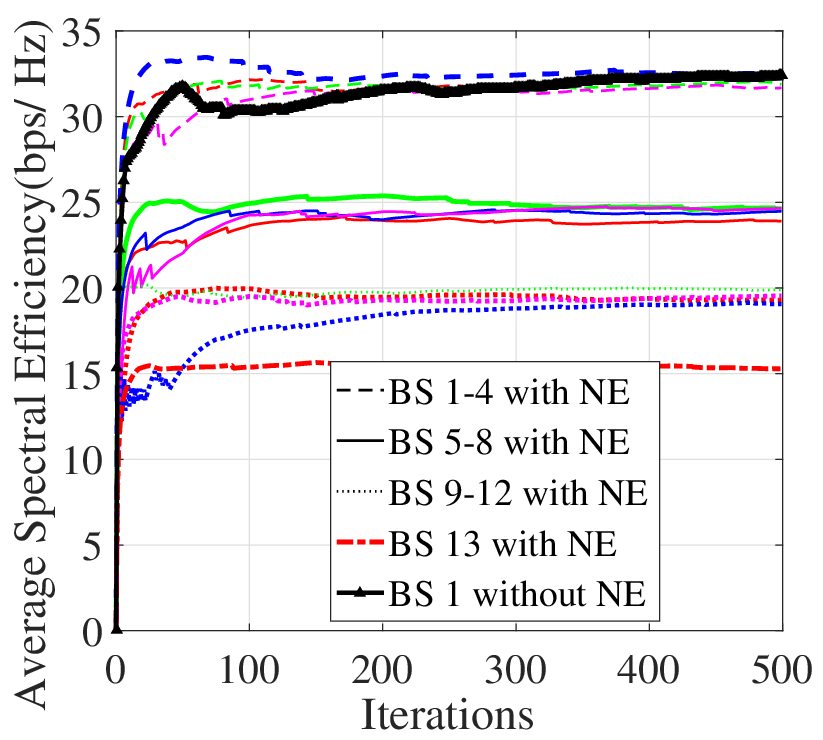}}
  \caption{Convergence of Algorithm \ref{alg_qlearning} for $K$=13 BSs with 32 TD sub-channels. }
  \label{fig_erw} 
   \vspace{-1.001em}
\end{figure}

Fig. \ref{fig_erw} shows the variations in network throughput for 13 cells as iterations progress. The black thick solid lines indicate  the maximum throughput potential of cell 1, assuming no interference from other BS signals. In Fig. \ref{fig_erw:32}, we see that 13 thin lines are convergent after 150 iterations. While in Fig. \ref{fig_erw:1},  they converge after 300 iterations. Although the thin lines approach their maximum values,  they remain  close to the thick line at convergence. This phenomenon implies that the proposed algorithm can obtain the optimal solutions, i.e., NE solutions.  In addition, the throughput of cells 9-13 is much smaller than that of cell 1--3 since the east cells are deeper than west cells.

Moreover, after comparing Fig. \ref{fig_erw:32} and Fig. \ref{fig_erw:1}, we find that the convergence performance of Algorithm \ref{alg_qlearning} with $N$=32 outperforms that using fair polling scheme. Also, the network for $N$=32 can  achieve a slightly higher NE throughput than the network using fair polling scheme.  These observations mainly stem from the fact that the agent consistently  selects the NE power strategy achieving the maximal Q value.   At $t$, the agent always chooses the VU at the  location with the strongest signals.  However, VUs with weak signals have a risk of  enduring long periods without service. In contrast, under the fair polling scheme, although the agent can only select one VU at $t$,  every VU gets serviced  within a TDMA superframe.

\vspace{-0.0em}
\section{Conclusion}\label{sect conclusions}
By modeling an antenna vibration system,  we derive the expressions of the  statistical characteristics (CDF, PDF and expectation) of the fast-fading channel for the CVMC. These derivations are validated by the Monte-Carlo's simulations. From these expressions and simulations, we find that the fast-fading channel has a remarkable effect on the VMIC.  More importantly,  the channel coefficients of the VMIC around the base station may be more even. Such evenly distributed coefficients may be beneficial for communication. Subsequently, we derive the expression of the CVMC outage probability and discuss its effects on the throughput of CMVC. They are also validated by the Monte-Carlo's simulations.   Also, from such expression, we find that the larger average AVI may not necessarily  reduce the  outage probability. In a few cases, it might greatly improve the outage performance according to our simulations. Finally,  we formulate a real-time power optimization problem for the multiple  competing  CVMC sub-networks.   To solve this problem, we propose the CVMC-PCO algorithm with NE strategy and multiagent Q-learning. Simulations show that the algorithm can quickly converge to the NE points. Based on the algorithm and the convergent simulations, two TDMA strategies are discussed. In the future studies such as cooperative MIC, special diversity gain and cross-layer protocols, the statistical characteristics of fast-fading channel of the CVMC  remain critical.
\vspace{-0.0em}


%

\appendices

\section{Proof of Lemma \ref{prop_jbiconvex}}\label{sect_proofjbiconvex}

The first derivative  of the  $J_{b,i+}$ and  $J_{b,i-}$ with respect to $X_i$ are
\begin{equation}\label{eqn_dJbi}
\begin{aligned}
\frac{\partial J_{b,i+}}{\partial X_i} &= \frac{(X_i-\frac{1}{2})\sin2\phi_i}{\sqrt{X_i-X_i^2}}-\cos2\phi_i,\\
\frac{\partial J_{b,i-}}{\partial X_i} &= -\frac{(X_i-\frac{1}{2})\sin2\phi_i}{\sqrt{X_i-X_i^2}}+\cos2\phi_i,
\end{aligned}
\end{equation}
respectively. According to~\eqref{eqn_dJbi}, the second derivative of $J_{b,i+}$ and $J_{b,i-}$  with respect to $X_i$  are
\begin{equation}\label{eqn_ddJbi}
\begin{aligned}
\frac{\partial^2 J_{b,i+}}{\partial X^2_i} &= \frac{\sin\! \left(2\phi_i\right)}{4\left(1-X_i\right)^{\frac{3}{2}}X_i^{\frac{3}{2}}},\\
\frac{\partial^2 J_{b,i-}}{\partial X^2_i} &= -\frac{\sin\! \left(2\phi_i\right)}{4\left(1-X_i\right)^{\frac{3}{2}}X_i^{\frac{3}{2}}},
\end{aligned} \notag
\end{equation}
respectively. For $X_i \in [0,1]$, when $2\phi_i \in [0, \pi\!+\!2k\pi]$ (i.e., $\phi_i \in [0, \frac{\pi}{2}\!+\!k\pi]$ and $\sin2\phi_i \geq 0$), it is evident that $\frac{\partial^2 J_{b,i+}}{\partial X^2_i}\geq 0$  and  $\frac{\partial^2 J_{b,i-}}{\partial X^2_i}\leq 0$ hold. Therefore, $J_{b,i+}(X_i)$ is convex and $J_{b,i-}(X_i)$ is concave. Similarly, when $\phi_i \in [- \frac{\pi}{2}\!-\!k\pi]$ and $\sin2\phi_i \leq 0$ which means  $\frac{\partial^2 J_{b,i+}}{\partial X^2_i}\leq 0$  and  $\frac{\partial^2 J_{b,i-}}{\partial X^2_i}\geq 0$,  $J_{b,i+}(X_i)$ is concave and $J_{b,i-}(X_i)$ is convex.

This concludes the proof.

\section{Proof of Theorem \ref{prop_jbicdf}}\label{sect_proofjbicdf}

Firstly, we derive the  CDF $F_{J_{b,i}}(z)$ of  $J_{b,i}$.

The CDF $F_{J_{b,i}}(z)$ of  $J_{b,i}$ can be obtained by  $F_{J_{b,i}}(z) = \mathbb{P}[J^{\sqrt{.}}_{b,i}(X_i^2) < z]$. However,  $F_{J_{b,i}}(z)$ is not continuous due to existence of the boundary $X_i=1$. We treat $X_i$ at the boundary point as a discrete random variable.

Since~\eqref{eqn_Jbi} is the conjugate pseudo-piecewise function, $F_{J_{b,i}}(z) = \mathbb{P}[J^{\sqrt{.}}_{b,i}(X_i) < z]$ can be further deduced into
\begin{equation}\label{eqn_CDFJpJn0}
\begin{aligned}
F_{J_{b,i}}(z) &= \mathbb{P}[J_{b,i+}(X_i | 0 \leq Y_i \leq 1) < z ] \\
               &\ \    +  \mathbb{P}[J_{b,i-}(X_i | -1 \leq Y_i \leq 0 ) < z].
\end{aligned}\notag
\end{equation}

Considering the even function $f_{Y_i}(y)$, we obtain
\begin{equation}\label{eqn_CDFJpJn}
\begin{aligned}
F_{J_{b,i}}(z) &= \frac{1}{2}\mathbb{P}[J_{b,i+}(X_i) \!<\! z ] \!+\! \frac{1}{2} \mathbb{P}[J_{b,i-}(X_i ) \!<\! z] \\
               &= \frac{1}{2}\left(F_{J_{b,i+}}(z)+ F_{J_{b,i-}}(z)\right).
\end{aligned}
\end{equation}

The solutions of inequalities  $J_{b,i+}(X_i, \phi_i) \!<\! z$ and  $J_{b,i-}(X_i, \phi_i) \!<\! z$  about $X_i$ depend on the  concavities of the $J_{b,i+}$ and $J_{b,i-}$, respectively, which   also depend on the sign of $\phi_i$. Fortunately, it is observed from Lemma \ref{prop_jbiXeven} that  solutions of  $J_{b,i+}(X_i, \phi_i) \!<\! z$ and  $J_{b,i-}(X_i, \phi_i) \!<\! z$ equal to  solutions of $J_{b,i-}(X_i, k\pi-\phi_i) \!<\! z$ and  $J_{b,i+}(X_i, k\pi-\phi_i) \!<\! z$. Therefore, without loss of generality, we can assume that $\phi_i \in [0,\frac{\pi}{2}]$ for both $F_{J_{b,i+}}(z)$ and $F_{J_{b,i-}}(z)$.

With the increase of $z$ and based on the continuity of $J_{b,i}$, we consider three cases of $z$ which are the cases of $z$$\in$$[0,\min\{\cos^2\phi_i, \sin^2\phi_i\})$, $z$$\in$$[\min\{\cos^2\phi_i, \sin^2\phi_i\},\max\{\cos^2\phi_i, \sin^2\phi_i\})$ and $z$$\in$$[\max\{\cos^2\phi_i, \sin^2\phi_i\}, 0)$  (i.e., [O,Z1), [Z1,Z2) and [Z2,Z3) in Fig. \ref{fig_JbiZX}), detailed in below.

\romannumeral1)  When $z\in [0, $Z1$]$,   $J_{b,i+}(x)=z$ has two solutions: $X_{i\mathrm{L}}$  and $X_{i\mathrm{H}}$. Thus, the solution for $J_{b,i+}(X_i) < z$ is $X_{i\mathrm{L}} < X_i < X_{i\mathrm{H}}$.
As a result, we get
 \begin{equation}
 \begin{aligned}
 F_{J_{b,i+}}(z) &= \mathbb{P}[X_{i\mathrm{L}}\! < \!X_i \!<\! X_{i\mathrm{H}}]
              \!=\!  \int^{X_{i\mathrm{H}}(z)}_{X_{i\mathrm{L}}(z)}p(x)dx.
 \end{aligned}\notag
 \end{equation}
Also when $z \in [0, {\rm Z1}]$, since the function $J_{b,i-}(X_i)$ is concave,  $J_{b,i-}$ is always greater than  $\min\{\cos^2\phi_i, \sin^2\phi_i\}$. This  indicates   $\mathbb{P}[J_{b,i-} < z] = 0$. Thus, Eq. \eqref{eqn_jbicdf:e} holds. 

\romannumeral2) We focus on the case where $z \in [$Z1$, $Z2$\}]$. Since the boundary AVI $X_i\!=\!1$ treated as a discrete variable,  we need to consider two sub-cases. For the sub-case where $\sin^2\phi_i\leq\cos^2\phi_i$,  the solution for $J_{b,i+}(X_i)<z$  is $x_{i\mathrm{L}} < X_i \leq 1$ which includes the boundary $X_i=1$. Therefore, we get
 \begin{equation}\label{eqn_jbipcdf4}
 \begin{aligned}
 F_{J_{b,i+}}(z) &= \mathbb{P}[X_{i\mathrm{L}} < X_i < 1] + \mathbb{P}[X_i\!=\!1] \\
              &=  \int^{1}_{X_{i\mathrm{L}}(z)}p(x)dx +  \int^{\infty}_{1}p(x)dx  \\
              &= 1 - \int_{1}^{X_{i\mathrm{L}}(z)}p(x)dx.
 \end{aligned}
 \end{equation}
Correspondingly, for $J_{b,i-}$, the solution for $J_{b,i-}(X_i)<z$  is $X_{i\mathrm{H}} < X_i \leq 1$ which also includes the boundary $X_i$=$1$.  we have
 \begin{equation}\label{eqn_jbincdf4}
 \begin{aligned}
 F_{J_{b,i-}}(z) &= \mathbb{P}[X_{i\mathrm{H}} < X_i < 1] + \mathbb{P}[X_i\!=\!1] \\
              &= 1 - \int_{1}^{X_{i\mathrm{H}}(z)}p(x)dx.
 \end{aligned}
 \end{equation}
 After substituting~\eqref{eqn_jbipcdf4} and~\eqref{eqn_jbincdf4} into~\eqref{eqn_CDFJpJn}, we obtain Eq. \eqref{eqn_jbicdf:d} in Theorem \ref{prop_jbicdf}.
 
     \begin{figure}[t]
 	\centering
 	\includegraphics[width=2.4in, height=1.8in]{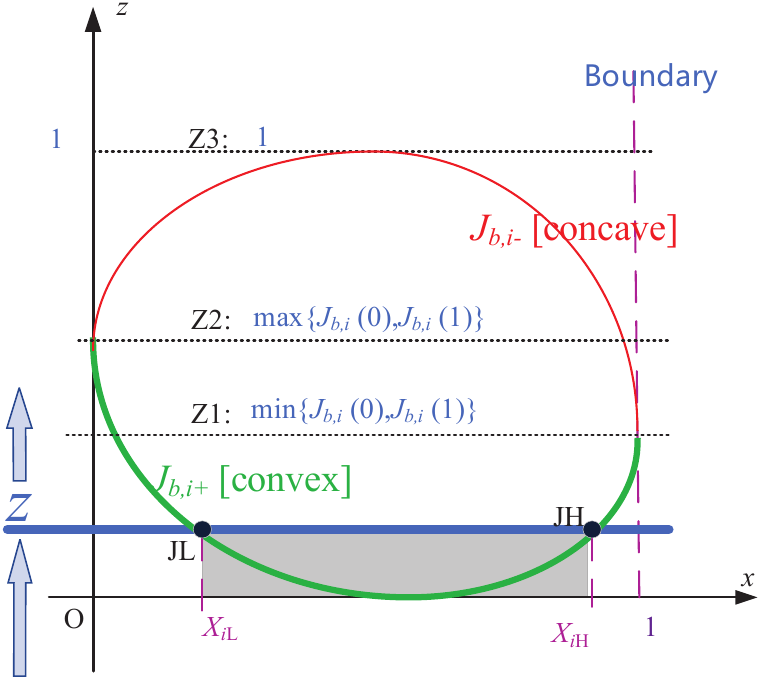}
 	\vspace{-0.5em}
 	\caption{ Cases of the inequality $J^{\sqrt{.}}_{b,i}(x) < z$ solutions.
 	}\label{fig_JbiZX}
 	\vspace{-1.0em}
 \end{figure}

 Similarly, for the sub-case where $\sin^2\phi_i\geq\cos^2\phi_i$,  the solutions of  $J_{b,i+}(X_i)<z$  and  $J_{b,i-}(X_i)<z$ are  $0 <X_i < X_{i\mathrm{H}}$ and   $0 <X_i < X_{i\mathrm{L}}$, respectively.  Furthermore, there are no boundary AVIs $X_i$  in these solutions. As a result, we get
  \begin{equation}\label{eqn_jbipcdf3}
 \begin{aligned}
 F_{J_{b,i+}}(z) &= \mathbb{P}[ 0 <X_i < X_{i\mathrm{H}}]
              = \int_{0}^{X_{i\mathrm{L}}(z)}p(x)dx,
 \end{aligned}
 \end{equation}
 and
  \begin{equation}\label{eqn_jbincdf3}
 \begin{aligned}
 F_{J_{b,i-}}(z) &\!=\! \mathbb{P}[ 0\! <\!X_i\! <\! X_{i\mathrm{L}}]
              \!=\!  \int_{0}^{X_{i\mathrm{H}}(z)}p(x)dx,
 \end{aligned}
 \end{equation}
 respectively.
  After substituting~\eqref{eqn_jbipcdf3} and~\eqref{eqn_jbincdf3} into~\eqref{eqn_CDFJpJn}, we obtain Eq. \eqref{eqn_jbicdf:c} in Theorem \ref{prop_jbicdf}.

\romannumeral3)  For the case of   $[$Z2$, $Z3$]$, all the values of  $J_{b,i+}(X_i)$ are smaller than $z\in [$Z2$, $Z3$\}]$, i.e.,
  \begin{equation}\label{eqn_jbipcdf2}
 \begin{aligned}
 F_{J_{b,i+}} &= 1. \\
 \end{aligned}
 \end{equation}
 Since  the function $J_{b,i-}(X_i)$ is concave, the solution of  $J_{b,i-}(X_i)<z$ is $0 \leq X_i < X_{i\mathrm{L}} \bigcup  X_{i\mathrm{H}} < X_i \leq 1 $ which includes the boundary $X_i=1$.
Then, we derive
   \begin{equation}\label{eqn_jbincdf2}
 \begin{aligned}
 F_{J_{b,i-}} &=  \mathbb{P}[J_{b,i-}(X_i)<z] \\
  &=  \mathbb{P}[ 0 \!<\!X_i \!<\! X_{i\mathrm{L}}] \!+\!  \mathbb{P}[  X_{i\mathrm{H}}< X_i \! <\! 1] \!+\!  \mathbb{P}[X_i \!=\! 1]\\
  &=   \int_{0}^{X_{i\mathrm{L}}(z)}p(x)dx \!+\! \int^{1}_{X_{i\mathrm{H}}(z)}p(x)dx \!+\!\int^{\infty}_{1}p(x)dx \\
  &= 1 - \int^{i\mathrm{H}(z)}_{X_{i\mathrm{L}}(z)}p(x)dx.
 \end{aligned}
 \end{equation}
  After substituting~\eqref{eqn_jbipcdf2} and~\eqref{eqn_jbincdf2} into~\eqref{eqn_CDFJpJn}, we have \eqref{eqn_jbicdf:b} in Theorem \ref{prop_jbicdf}.

  For  $J_{b,i} =\cos^2(\theta_i+\phi_i) \in [0,1]$, both~\eqref{eqn_jbicdf:a} and~\eqref{eqn_jbicdf:f} in Theorem \ref{prop_jbicdf} are always valid.

   To sum up, the CDF of the $J_{b,i}$ can be expressed by~\eqref{eqn_jbicdf:a}--\eqref{eqn_jbicdf:f} in Theorem \ref{prop_jbicdf}.

  Secondly, we discuss the PDF of the $J_{b,i}$ as $z$ increases from 0 to 1 which involves six cases as below.

\uppercase\expandafter{\romannumeral1})  When $z\in [0, $Z1$\})$, $F_{J_{b,i}}$ is continuous and differentiable. Thus, we get
  \begin{equation}\label{eqn_jbipdf10}
  \begin{aligned}
  f_{J_{b,i}}(z) &=  \left(\frac{\int^{X_{i\mathrm{H}}(z)}_{X_{i\mathrm{L}}(z)}p(x)dx}{2}\right)' \\
           &=   \frac{ X^{'}_{i\mathrm{H}}(z)  p(X_{i\mathrm{L}}(z))-    X^{'}_{i\mathrm{L}}(z) p(X_{i\mathrm{L}}(z))}{2},
  \end{aligned}
  \end{equation}
where the derivatives of $X^{'}_{i\mathrm{H}}(z)$ and $X^{'}_{i\mathrm{L}}(z)$ are detailed in~\eqref{eqn_JbiInvDeriv}. Since $J_{b,i+}$ is convex, $X^{'}_{i\mathrm{H}}(z)$ and $X^{'}_{i\mathrm{L}}(z)$ are monotonically increasing and decreasing, respectively. Namely, $X^{'}_{i\mathrm{H}}(z)>0$ and $X^{'}_{i\mathrm{H}}(z)<0$. Eq. ~\eqref{eqn_jbipdf10} transforms into

  \begin{equation}\label{eqn_jbipdf1}
  \begin{aligned}
  f_{J_{b,i}}(z) \!= \!   \frac{| X^{'}_{i\mathrm{L}}(z) | p(X_{i\mathrm{L}}(z)) \!+ \!  | X^{'}_{i\mathrm{H}}(z)| p(X_{i\mathrm{L}}(z))}{2},
  \end{aligned}
  \end{equation}
which falls under~\eqref{eqn_jbipdf:a} in Theorem \ref{prop_jbicdf}.

\uppercase\expandafter{\romannumeral2})  When $z$$\in$[$Z1$, $Z2$) and if $J(1)$$\leq$$J(0)$,  $F_{j_{b,i}}(z)$ is  continuous and differentiable. Furthermore, both $X^{'}_{i\mathrm{H}}(z)$ and $X^{'}_{i\mathrm{L}}(z)$ are monotonically decreasing, i.e.,  $X^{'}_{i\mathrm{H}}(z)$$=$$ -|X^{'}_{i\mathrm{H}}(z)| $ and $X^{'}_{i\mathrm{L}}(z)$$=$$ -|X^{'}_{i\mathrm{L}}(z)| $.  Therefore, upon substitution of these two equations into $ \partial(1$$-$$\!\frac{1}{2}\int^{ X_{i\mathrm{L}}(z)}_{0}p(x)dx$$-$$\frac{1}{2}\int^{X_{i\mathrm{H}}(z)}_{0}p(x)dx)/\partial z $, we obtain the identical expression as~\eqref{eqn_jbipdf1}.

\uppercase\expandafter{\romannumeral3}) When $z \in ($Z1$, $Z2$)$ and if $J(1) \geq J(0)$,  $F_{J_{b,i}}(z)$ is  continuous and differentiable. Both $X^{'}_{i\mathrm{H}}(z)$ and $X^{'}_{i\mathrm{L}}(z)$ are monotonically increasing, i.e.,  $X^{'}_{i\mathrm{H}}(z)$$=$$|X^{'}_{i\mathrm{H}}(z)|$ and $X^{'}_{i\mathrm{L}}(z)$$=$$|X^{'}_{i\mathrm{L}}(z)|$.  Therefore, after substituting these two equations into $ \partial(\frac{1}{2}\int^{ X_{i\mathrm{L}}(z)}_{0}p(x)dx$$+$$\!\frac{1}{2}\int^{X_{i\mathrm{H}}(z)}_{0}p(x)dx)/\partial z$, we obtain the identical expression as~\eqref{eqn_jbipdf1}.

\uppercase\expandafter{\romannumeral4}) When $z \in ($Z2$, $Z3$)$,  $F_{j_{b,i}}(z)$ is  continuous and differentiable.  $X^{'}_{i\mathrm{H}}(z)$ and $X^{'}_{i\mathrm{L}}(z)$ are monotonically decreasing and increasing, respectively, i.e.,  $X^{'}_{i\mathrm{H}}(z)$$=$$-|X^{'}_{i\mathrm{H}}(z)| $ and $X^{'}_{i\mathrm{L}}(z)$$=$$|X^{'}_{i\mathrm{L}}(z)| $.  Therefore, after substituting these two equations into $ \partial(1$$-$$\frac{1}{2}\int^{ X_{i\mathrm{L}}(z)}_{0}p(x)dx $$-$$\frac{1}{2}\int^{X_{i\mathrm{H}}(z)}_{0}p(x)dx)/\partial z$, we obtain the identical expression as~\eqref{eqn_jbipdf1}.

\uppercase\expandafter{\romannumeral5}) When $z = J(0)=\cos^2\phi_i$,  $F_{j_{b,i}}(z)$ is  continuous at $z = J(0)$. According to cases \uppercase\expandafter{\romannumeral1}), \uppercase\expandafter{\romannumeral2}) \uppercase\expandafter{\romannumeral3})and \uppercase\expandafter{\romannumeral4}), the left derivative of $F_{J_{b,i}}(J(0))$  equals to its right derivative. Therefore, we obtain the same expression as~\eqref{eqn_jbipdf1}.

\uppercase\expandafter{\romannumeral6})   When $z = J(1)=\sin^2\phi_i$, there exist the Dirac's function in $f_{X_i}(x)$. That is to say,  $F_{j_{b,i}}(z)$ is not continuous at $z = J(1)$. To obtain $f_{j_{b,i}}(z)$,  we treat $J_{b,i} = J^{\sqrt{.}}_{b,i}(1)$ as the sum of a discrete and a continuous variable and then use the principle of the physical meaning of PDF.   For $z = J^{\sqrt{.}}_{b,i}(1) =\sin^2\phi_i$, it is obvious that  one of the solution of $J^{\sqrt{.}}_{b,i}(X_i) =\sin^2\phi_i$ is  $X_i = 1$.   Without loss of generality, we assume that another solution of  $J^{\sqrt{.}}_{b,i}(X_i) =\sin^2\phi_i$ is $X_{i} = X_{i{\mathrm{L}}}(\sin^2\phi_i)$.
 According to the physical meaning of PDF, the PDF of $J_{b,i}$ can be treated as
\begin{equation}\label{eqn_proofpdf2}
\begin{aligned}
  f_{J_{b,i}}(J^{\sqrt{.}}_{b,i}(1)) &= \lim\limits_{dz\rightarrow0}\bigg( \mathbb{P}[z \!=\! J^{\sqrt{.}}_{b,i}(1) | X_i = 1] \\
          &  \!+\!      \mathbb{P}[z \!=\! J^{\sqrt{.}}_{b,i}(1) | X_i \!=\! X_{i{\mathrm{L}}}(\sin^2\phi_i)] \bigg)/dz.
\end{aligned}
\end{equation}
Since $F_{X_i}(x)$ at $X_{i{\mathrm{L}}}(\sin^2\phi_i)$ is continuous and differentiable, $X_{i{\mathrm{L}}}(\sin^2\phi_i)$  is a continuous random variable.  This implies that $\mathbb{P}[z$$=$$ J^{\sqrt{.}}_{b,i}(1)$$|$$X_i$$=$$X_{i{\mathrm{L}}}(\sin^2\phi_i)]$$\rightarrow$$0$.  As derived in the description of Definition \ref{def_bpd}, $\lim\limits_{dz\rightarrow0}\frac{1}{dz}$=$\delta(0)$ holds. Since $F_{X_i}(x)$ at $X_i$$=$1 is discrete, $\mathbb{P}[z $$=$$ J^{\sqrt{.}}_{b,i}(1)$$|$$X_i$$=$$1]$=$\mathbb{P}[X_i$$=$$1]$$=$
$\int^{\infty}_{1}p(x)dx$ is satisfied.  Thus, Eq. \eqref{eqn_proofpdf2} transforms into~\eqref{eqn_jbipdf:b} in Theorem \ref{prop_jbicdf}.

 To sum up, the PDF of the $J_{b,i}$ can be expressed by~\eqref{eqn_jbipdf:a}--\eqref{eqn_jbipdf:c} in Theorem \ref{prop_jbicdf}.

This concludes the proof.

\section{Proof of Proposition \ref{prop_jbiej2normprop}}\label{sect_proofjbiej2normprop}

To get the monotonicity of  $\mathbb{E}(J_{b,i})$ with $\sigma^2_i$, we derive the derivative of $\mathbb{E}(J_{b,i})$  as
\begin{equation}\label{eqn_dEJbiNorm}
\begin{aligned}
\tfrac{\partial \mathbb{E}(J_{b,i})}{\partial (\sigma_i^2)} &= \left(\sqrt{\tfrac{2}{\sigma_i^2\pi}}\,{\mathrm e}^{-\tfrac{1}{2\sigma^2_i}}-\mathrm{erf}\! \left(\sqrt{\tfrac{1}{2\sigma_i^{2}}}\right)\right)\cos\! \left(2\phi_i\right).\notag
\end{aligned}
\end{equation}
It is obviously observed that
$\lim\limits_{\sigma_i \rightarrow \infty} \frac{\partial \mathbb{E}(J_{b,i})}{\partial (\sigma^2_i)} = 0$,
Consequently, we derive the second derivative of $\mathbb{E}(J_{b,i})$ as
\begin{equation}
\begin{aligned}
\frac{\partial^2 \mathbb{E}(J_{b,i})}{\partial (\sigma_i^2)^2} &= \frac{{\mathrm e}^{-\frac{1}{2\sigma_i^{2}}}}{\sqrt{2\pi\sigma_i^{10}}}  \cos\! \left(2\phi_i\right).\notag
\end{aligned}
\end{equation}
Since $\frac{{\mathrm e}^{-\frac{1}{2\sigma_i^{2}}}}{\sqrt{2\pi\sigma_i^{10}}}$ is always positive, the sign of $\frac{\partial^2 \mathbb{E}(J_{b,i})}{\partial (\sigma_i^2)^2}$ is determined by  $\cos\!\left(2\phi_i\right)$. That is to say, for $\phi_i\in[ k\pi,\frac{\pi}{4}+k\pi)$ and $ \cos\! \left(2\phi_i\right)\geq0$, the first derive $\frac{\partial \mathbb{E}(J_{b,i})}{\partial (\sigma_i^2)}$ is monotonically increasing and has  the upper bound 0, i.e.,  $\frac{\partial \mathbb{E}(J_{b,i})}{\partial (\sigma^2_i)} \leq  0$. Therefore,  $\mathbb{E}(J^{\sqrt{.}}_{b,i}(\sigma^2_i))$ is  monotonically decreasing when $\phi_i\in[ k\pi,\frac{\pi}{4}+k\pi)$. Conversely, $\mathbb{E}(J^{\sqrt{.}}_{b,i}(\sigma_i))$ is  monotonically increasing when  $\phi_i\in(\frac{\pi}{4}+k\pi, \frac{\pi}{2}+k\pi]$ with $\cos\! \left(2\phi_i\right) < 0$.

This concludes the proof.


\ifCLASSOPTIONcaptionsoff
  \newpage
\fi

\bibliographystyle{IEEEtran} 
\bibliography{IEEEabrv,MIref}

\vspace{6pt}

\end{document}